\documentclass{aa}

\usepackage{amssymb}
\usepackage{amsmath}
\usepackage{graphicx}
\usepackage{natbib}
\usepackage{multirow}
\usepackage{txfonts}
\usepackage{microtype}
\usepackage{color}
\usepackage[colorlinks=true,citecolor=blue,linkcolor=blue]{hyperref}
\usepackage{epstopdf}
\usepackage[flushleft]{threeparttable}
\usepackage{booktabs}

\begin{document}

\title{The broad-line type Ic SN 2020bvc: signatures of an off-axis gamma-ray burst afterglow}
\titlerunning{SN 2020bvc}
\authorrunning{Izzo et al.}
\author{L. Izzo$^{1}$, K. Auchettl$^{1,2,3,4}$, J. Hjorth$^{1}$, F. De Colle$^{5}$, C. Gall$^{1}$, C. R. Angus$^{1}$, S. I. Raimundo$^{1}$,  E. Ramirez-Ruiz$^{1,3}$}
\institute{$^{1}$ DARK, Niels Bohr Institute, University of Copenhagen, Lyngbyvej 2, DK-2100 Copenhagen, Denmark\\
$^{2}$ School of Physics, The University of Melbourne, Parkville, VIC 3010, Australia\\ $^{3}$ARC Centre of Excellence for All Sky Astrophysics in 3 Dimensions (ASTRO 3D), Australia\\
$^{4}$ Department of Astronomy and Astrophysics, University of California, Santa Cruz, CA 95064, USA\\
$^{5}$ Instituto de Ciencias Nucleares, Universidad Nacional Aut\'onoma de M\'exico, Apartado Postal 70-543, 04510 México, CDMX\\ 
}

\abstract
{Long-duration gamma-ray bursts (GRBs) are almost unequivocally associated with very energetic, broad-lined supernovae (SNe) of Type Ic-BL. While the gamma-ray emission is emitted in narrow jets, the SN emits radiation isotropically. Therefore, some SN Ic-BL not associated with GRBs have been hypothesized to arise from events with inner engines such as off-axis GRBs or choked jets. Here we present observations of the nearby ($d = 120$ Mpc) SN 2020bvc (ASAS-SN 20bs) which support this scenario. \textit{Swift} UVOT observations reveal an early decline (up to two days after explosion) while optical spectra classify it as a SN Ic-BL with very high expansion velocities ($\approx$ 70,000 km/s), similar to that found for the jet-cocoon emission in SN 2017iuk associated with GRB 171205A. Moreover, \textit{Swift} X-Ray Telescope and \textit{CXO} X-ray Observatory detected X-ray emission only three days after the SN and decaying onwards, which can be ascribed to an afterglow component. Cocoon and X-ray emission are both signatures of jet-powered GRBs. In the case of SN 2020bvc, we find that the jet is off axis (by $\approx$ 23 degrees), as also indicated by the lack of early ($\approx 1$ day) X-ray emission which explains why no coincident GRB was detected promptly or in archival data. These observations suggest that SN 2020bvc is the first orphan GRB detected through its associated SN emission.
}


\date{}

\maketitle

\section{Introduction}\label{sec:1}
Type Ic broad-lined (BL) supernovae (SN) form a particular class of core-collapse SNe characterized by the presence of broad absorption features (full-width half-maximum 
$\approx$ 10,000 km/s) and extreme expansion velocities ($v_{\rm exp} \sim$ 20,000 km/s) at the maximum of the SN emission \citep{Modjaz2016,Gal-Yam2017}. The lack of hydrogen and helium in their spectra imply that their progenitors are compact massive stars, likely Wolf-Rayet stars with initial masses M $\approx$ 25--30 M$_{\odot}$  \citep{Woosley2006} and generally characterized by low-metallicity abundances \citep{Sanders2012,Taddia2019}. SNe Ic-BL are also the only flavour of SNe connected with gamma-ray bursts (GRB, \citealp{Galama1998,Hjorth2003,2003ApJ...591L..17S, Kaneko2007}), 
although the fraction of SNe Ic-BL associated with GRBs is only $\approx 10 \%$ of the entire population of SNe Ic-BL \citep{Soderberg2006,Guetta2007,Cano2017}. 
Lack of GRB emission may be due 
to the presence of jet emission that is not aligned with our line-of-sight,
given that GRB jets have average collimation angles of $\theta \lesssim 10$ deg \citep{Frail2001,Kumar2015}. Another possibility is that the jet is not sufficiently fed by the inner central engine such that it fails to break out of the progenitor star \citep[the ''choked-jet''scenario,][]{MacFadyen2001,Ramirez-Ruiz2002, Lazzati2012}. Extended surveys at radio frequencies have excluded the presence of off-axis jet emission in samples of SNe Ic-BL without an associated GRB \citep{Soderberg2006,Corsi2016}, although more recently, for the case of the Ic-BL SN 2014ad \citep{Stevance2017,Sahu2018}, an off-axis angle of $\theta \gtrsim 30$ deg could not be ruled out \citep{Marongiu2019} 
but see also \citet{Ho2019}.

\begin{figure*}[h!]
\centering
\includegraphics[width=0.98\hsize,clip]{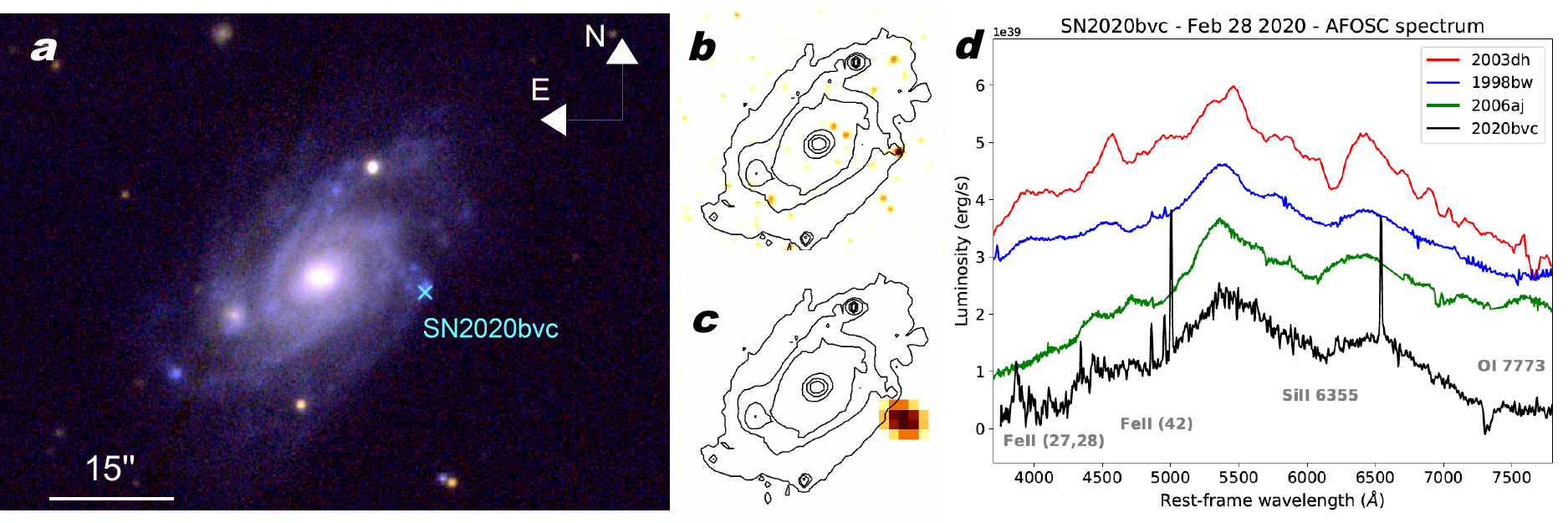}
\caption{ {\it (a)} The host galaxy of SN 2020bvc, UGC 9379, as imaged by the Pan-STARRS PS1 telescope. The image was created using the $g'$, $i'$ and the $z'$ images as single channels of an RGB image. The cyan cross marks the position of SN 2020bvc, offset 18\arcsec\ from the nucleus its host galaxy. 
{\it (b, c)} {\emph CXO} (upper panel) and {\emph Swift}-XRT (lower panel) count maps image superimposed to the contours of the PS1 $i$-band archival image of UGC 9379. The {\emph CXO} image has better spatial resolution, showing the proximity of SN 2020bvc to the \ion{H}{ii} region. {\it (d)} Spectrum of SN 2020bvc obtained with the AFOSC spectrograph at the Cima Ekar observatory in Asiago, Italy (black curve) on February 28, 2020, i.e., 24 days after the discovery of the SN, and compared with the spectra of three GRB SNe (arbitrarily offset) at similar epochs: SN 1998bw (blue -- day 28, \citealp{Galama1998}), SN 2003dh (red -- 28 days, \citealp{Hjorth2003}) and SN 2006aj (green -- day 19, \citealp{Sollerman2006}).}
\label{fig:4}
\end{figure*}

In the standard GRB scenario, the jet is highly collimated when it breaks out of the progenitor star. The jet opening angle is expected to be inversely related to the bulk Lorentz factor, $\Gamma$, of the expanding radiating plasma. This has been confirmed in theoretical \citep{Tchekhovskoy2009,Komissarov2010} and population synthesis simulations \citep{Ghirlanda2013}. However, as the jet expands in the circumburst medium, it decelerates due to the interaction with the medium, 
implying a decrease of $\Gamma$ and a consequent increase of the jet beaming angle \citep{Rhoads1997}. Depending on the initial direction of the jet axis, the beaming angle of the jet will reach our line of sight at later times, which is when the afterglow starts to become visible as weak and soft X-ray emission (the ''orphan'' GRB afterglow, \citealp{Granot2002,Kumar2003,Piran2005}). The early interaction of the jet with the dense stellar layers of the progenitor star also gives rise to an expanding ''{\it cocoon}'' of material that spreads laterally with respect to the direction of the collimated jet emission \citep{Meszaros2001}. 
The jet deposits a considerable amount of kinetic energy into the cocoon, which expands with mildly-relativistic velocities once it breaks out of the progenitor star \citep{Ramirez-Ruiz2002,Bromberg2011b}. A choked jet, which fails to escape its progenitor star, transfers all its energy into the cocoon component \citep{Bromberg2011,Bromberg2012,Irwin2019}. 
The cocoon emission is expected to be detectable a few hours after the collapse of the progenitor star \citep{Ramirez-Ruiz2002} and, once it breaks out, it starts to expand into the interstellar medium. The main observational signatures for this emission are rapid ($t \approx$ 2--3 days) cooling, hot thermal emission and the presence of high-velocity absorption features of Fe-peak elements in the optical spectra  \citep{Nakar2017,Izzo2019}. 
A considerable fraction of "engine-driven" SNe Ic-BL 
could have relativistic ejecta and could be detected at frequencies other than gamma-rays. In recent years, some SNe Ic-BL have exhibited bright radio emission that can be explained with mildly-relativistic velocity ejecta interacting with the circumburst medium \citep{Soderberg2010,Margutti2014,Milisavljevic2015,Chakraborti2015}. Similarly, the detection of young SNe Ic-BL can also be used to pinpoint orphan GRB afterglows, in particular at X-ray energies.

In this Letter we present the first case of an off-axis GRB discovered via its associated SN.\footnote{For GRB 020410, \citet{Levan2005} used late-time (28 days) {\it Hubble Space Telescope} observations of a SN lightcurve bump to identify the GRB optical afterglow in early optical data. Similarly, for FIRST J141918.9+394036, \citet{Marcote2019} presented evidence for an off-axis afterglow on the basis of the radio emission alone.} SN 2020bvc (or ASASSN-20bs) is a type Ic-BL SN recently discovered by the ASAS-SN survey \citep{2014ApJ...788...48S} in the nearby ($d = 120$ Mpc) galaxy UGC 9379. SN 2020bvc was discovered on February 4.6 2020 UT as a blue ($g' = 17$ mag) rising source \citep{Stanek2020}. The most stringent upper limit ($g' = 18.6$ mag) provided by the ASAS-SN survey was obtained on February 3.6 UT, one day prior to the discovery. Spectral observations taken with the FLOYDS spectrograph at the Faulkes-North Telescope on February 5.54 UT suggested a blue featureless continuum \citep{Hiramatsu2020} while a subsequent spectrum taken with SPRAT at the Liverpool Telescope on February 8.24 UT confirmed the type Ic-BL nature of SN 2020bvc by matching the observed spectrum with SN 1998bw six days before peak brightness \citep{Perley2020}. SN 2020bvc was detected at X-ray frequencies with the \textit{Chandra X-ray Observatory (CXO)} \citep{Ho2020a} and with the Neil Gehrels \textit{Swift} Observatory (target ID = 032818, Observations 12-20). SN 2020bvc was also observed at radio frequencies ($\nu = 10$ GHz, $F_{\nu} = 63\pm6 \mu$Jy) with the Very Large Array (VLA) \citep{Ho2020b} on February 16.67 UT.

\section{Observations}\label{sec:2}

We observed SN 2020bvc with the AFOSC spectrograph, mounted on the 1.82-meter telescope of the Asiago Cima Ekar Observatory on February 28.95 UT. A series of 3$\times$900 s spectra was obtained for the SN and a single spectrum for a spectrophotometric standard for the flux calibration. We reduced the data using the standard {\tt IRAF} \citep{Tody1986} procedure for long-slit spectra. The final calibrated spectrum is shown in Figure \ref{fig:4}, where we show the comparison with GRB-SNe 1998bw, 2003dh and 2006aj at similar epochs. The strong similarity between them confirms the type Ic-BL nature of SN 2020bvc. The position of SN 2020bvc is very close to a bright \ion{H}{ii} region of its host galaxy, UGC 9379 (Fig. \ref{fig:4}).

\begin{figure}
\centering
\includegraphics[width=0.95\hsize,height=0.27\vsize,clip]{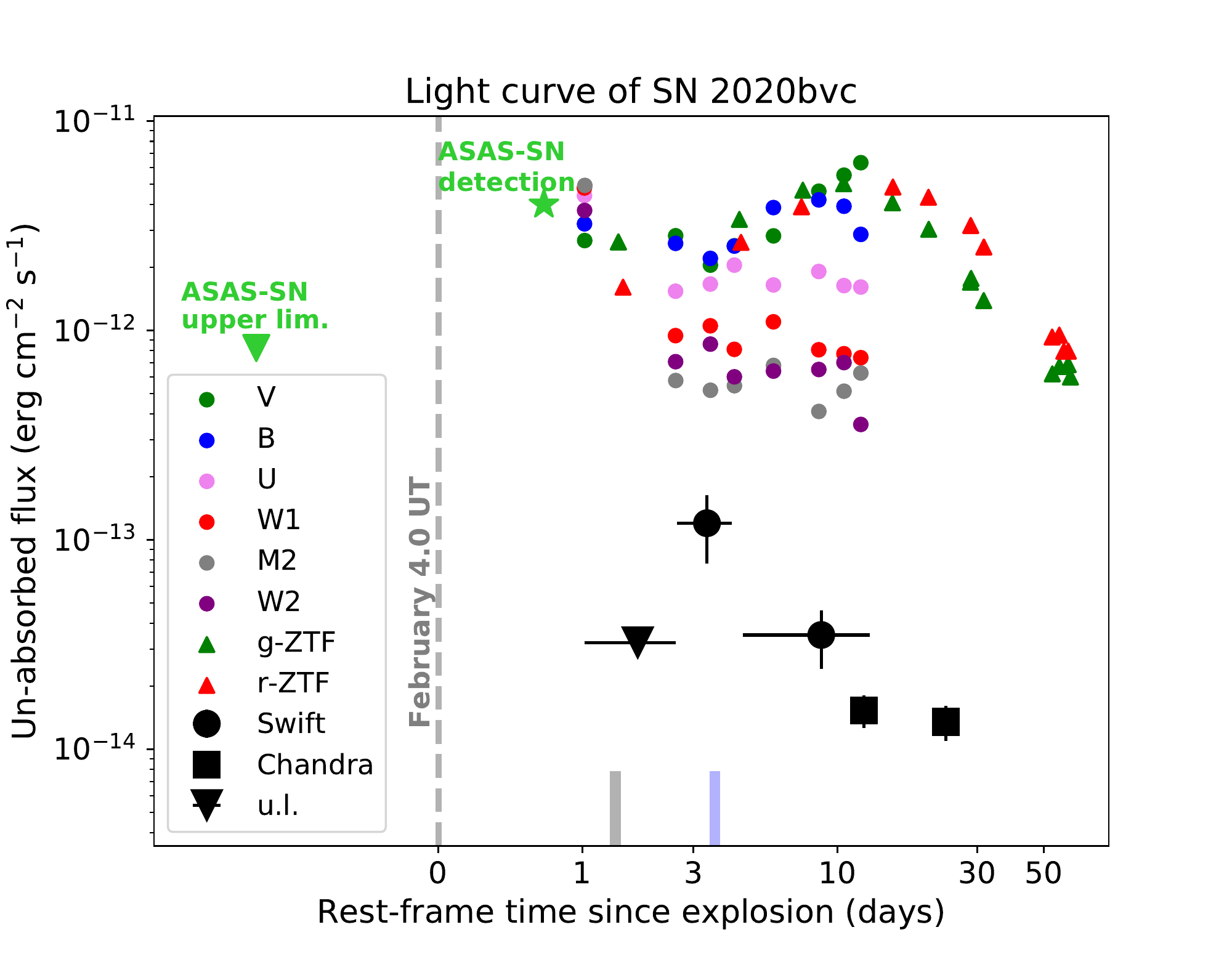}
\caption{The optical and X-ray light curve of SN 2020bvc. X-ray (\textit{Swift}-XRT and \textit{CXO}, 0.3--10 keV) data are plotted in black while \textit{Swift} UVOT data are represented with colored circles (Tab. \ref{tab:App1}). Zwicky Transient Facility (ZTF, \citealp{Graham2019,Bellm2019}) $g$ and $r$ data (Tab. \ref{tab:App2}) are shown as triangles. We also indicate the detection and the last non-detection by the ASAS-SN survey. Time is relative to the estimated epoch of SN explosion (see Sec. \ref{sec:4}), marked with a gray dashed line. 
The epochs of the FLOYDS and SPRAT spectra are marked in gray and blue.}
\label{fig:1}
\end{figure}

We extracted the \textit{Swift} optical/UV light curves using a 5 arcsec aperture and the \textit{Swift} analysis program \textsc{uvotsource}. We used a sky region of 20 arcsec radius to estimate and subtract the sky background. The UVOT magnitudes were derived assuming the \emph{most up to date} UVOT calibration \citep{2010MNRAS.406.1687B}. We did not attempt host galaxy subtraction for the \textit{Swift} light curves, but the UVOT data were corrected for the Galactic extinction in the direction of SN 2020bvc, E(B$-$V) = 0.01 mag \citep{Schlafly2011}, using a \citet{Fitzpatrick1999} dust extinction function. 
Simultaneously, SN 2020bvc was observed using the X-Ray Telescope (XRT) onboard \textit{Swift}. All observations were analyzed using {\tt xrtpipeline} version 0.13.2, the standard filters and screening as suggested by the \textit{Swift} data reduction guide\footnote{\url{https://swift.gsfc.nasa.gov/analysis/xrt_swguide_v1_2.pdf}} and the \emph{most up to date} CALDB. To place constraints on the presence of X-ray emission, we used a source region with a radius of 10 arcsec centered on the position of SN 2020bvc\footnote{A 10 arcsec radius corresponds to an encircled energy fraction of $\sim$55\% at 1.5 keV assuming on-axis pointing \citep{2004SPIE.5165..232M}.} and a source-free background region with a radius of 75 arcsec located at RA = 14:34:07.0, Dec =$+$40:13:57.6 (J2000). In addition to the \textit{Swift} UVOT/XRT observations we also analyzed two public Directory Discretionary Time \textit{CXO} observations (ObsID: 23171 and 23172; PI: Ho) of SN 2020bvc that were taken on February 16 and February 28. All data were analyzed using \textsc{CIAO} version 4.12 and the \emph{most up to date} CALDB. To extract a count rate from the \textit{CXO} data, we used a source region of 2 arcsec centered on SN 2020bvc and a 20 arcsec source free background region centered at RA = 14:33:59.1, Dec =$+$40:15:09.9 (J2000). All extracted count rates were corrected for encircled energy fraction\footnote{\url{https://cxc.harvard.edu/proposer/POG/html/chap4.html}}. To increase signal-to-noise of the \textit{Swift} XRT observations, we merged the first two, second two and last four observations using \textsc{xselect} version 2.4g. In all but the first epoch of the merged \textit{Swift} XRT observations, we detect faint X-ray emission arising from the position of SN 2020bvc (see Fig. \ref{fig:1}). To convert the extracted count rates (both the 3$\sigma$ upper limit and the detections), we assume an absorbed power law with a photon index of 2, a redshift $z = 0.02524$ of the host galaxy of SN 2020bvc and a Galactic column density of 9.9$\times10^{19}$ cm$^{2}$ along the line of sight \citep{2016A&A...594A.116H}. The UVOT and XRT light curves are shown in Figure \ref{fig:1}.

\begin{figure*}[h!]
\centering
\includegraphics[width=0.40\hsize,clip]{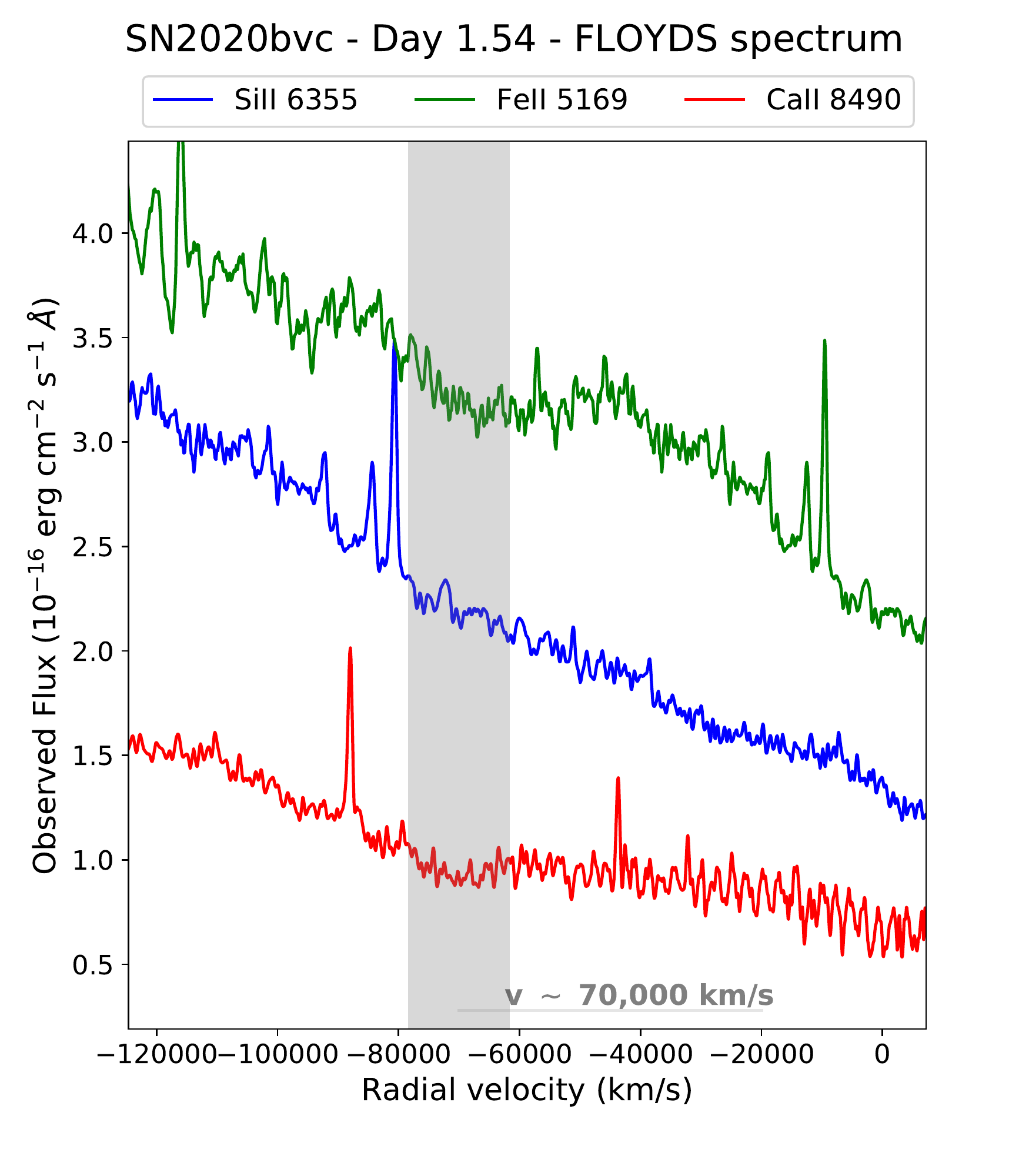}
\includegraphics[width=0.40\hsize,clip]{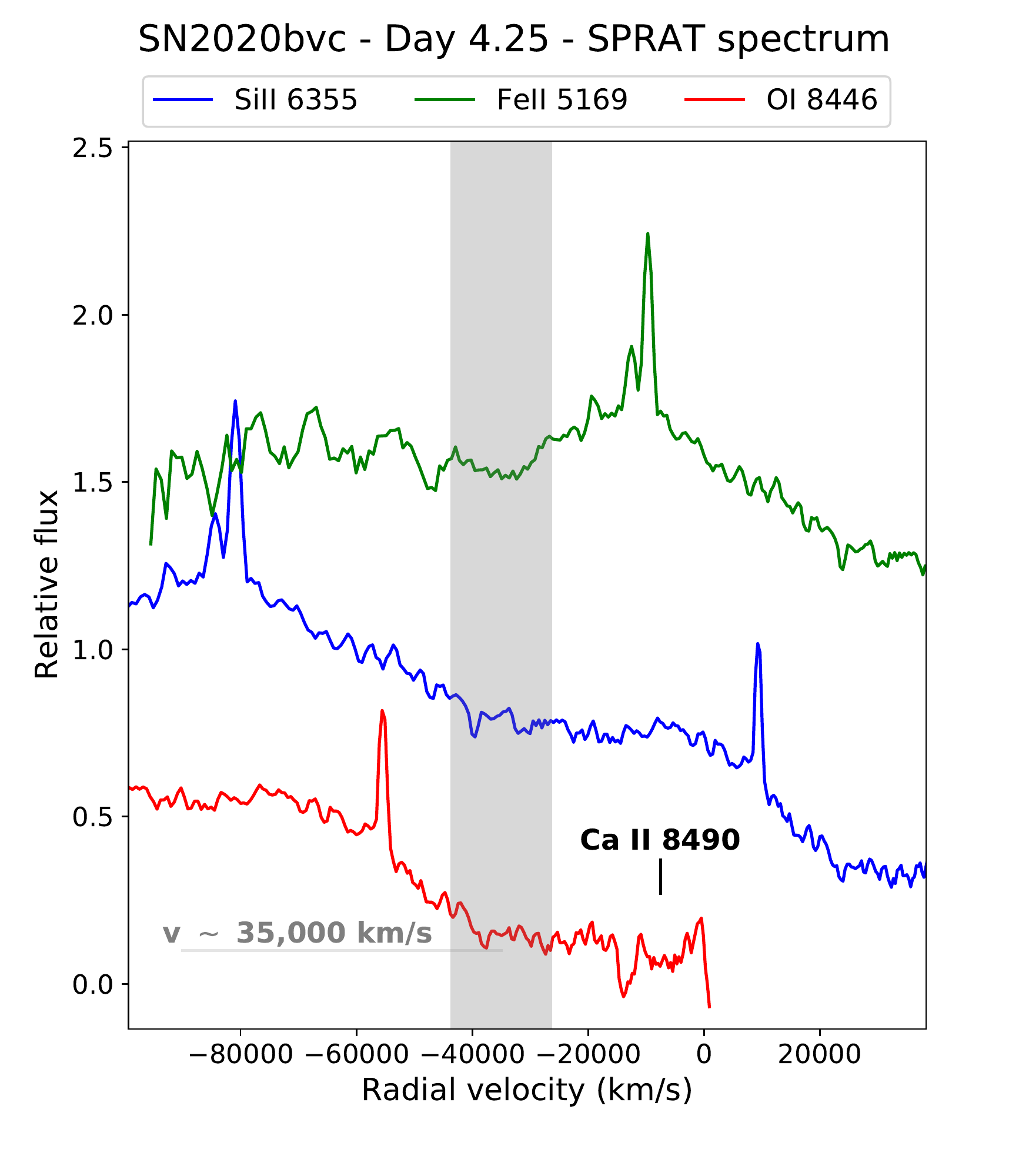}
\caption{{\it (Left panel)} The FLOYDS spectrum of SN 2020bvc observed one day after the SN discovery and centered on three different atomic transitions (\ion{Fe}{ii} 5169, \ion{Si}{ii} 6355 and \ion{Ca}{ii} 8490). The gray band marks the presence of an absorption feature at $v_{\rm exp} \approx$ 70,000 km/s, common to \ion{Fe}{ii} and \ion{Ca}{ii} while it is not observed for \ion{Si}{ii}. {\it (Right panel)} The SPRAT spectrum of SN 2020bvc observed three days after the SN discovery. In this panel the red spectrum is centered on the \ion{O}{i} 8446 line, with the position of the features indicated for a velocity of $v_{\rm exp} \approx$ 35,000 km/s, common to \ion{Fe}{ii} and \ion{O}{i} and barely discernible for the \ion{Si}{ii} line.}
\label{fig:3}
\end{figure*}

\section{Analysis of SN 2020bvc}\label{sec:3}

The UVOT light curve shows the presence of rapidly decaying UV/optical emission observed in the first three days since the SN explosion (
from here on, we fix our Day 0 to February 4.0 UT, see also Section 4) and is characterized by fast color evolution. We built two spectral energy distributions (SEDs) using the first two epochs of UVOT data (Day 1.0 and Day 4.3 after the SN explosion, respectively).  We found that these SEDs can be fitted with a thermal component with the black-body temperature varying from $T_{1.0} =$ 12,300 K to $T_{4.3} =$ 6,100 K (see the complete analysis in the Sec. \ref{app:2}). The radius of the thermal emitter rapidly evolves from $R_{1.0} =$ (1.42$\pm$0.56)$\times$10$^{10}$ km to $R_{4.3} =$ (4.43$\pm$1.59)$\times$10$^{10}$ km, similarly to the SN 2017iuk case \citep{Izzo2019}. 

We have also analyzed the FLOYDS and the SPRAT optical spectra, available on the WISeREP\footnote{\url{https://wiserep.weizmann.ac.il/}}. The FLOYDS spectrum reveals the presence of broad absorption centered at 4300 $\AA$ and 6900 $\AA$. We interpret these as the multiplet 42 of \ion{Fe}{ii} and the near-IR \ion{Ca}{ii} triplet at an expanding blueshifted velocity of $v_{\rm exp,1} \approx$ 70,000 km/s (see Figure \ref{fig:3}). These features remain visible in the later SPRAT spectrum but with a lower expansion velocity of $v_{\rm exp,2} \approx$ 35,000 km/s. The SPRAT spectrum also shows signatures of \ion{O}{i} 7775 $\AA$ and the possible presence of \ion{Si}{ii} 6355 $\AA$ (absent in the earlier FLOYDS spectrum) at similar velocities. 

Finally, from the analysis of the nebular emission lines visible in the Asiago spectrum, which originate from the gas surrounding the SN, we have inferred the physical properties of the gas itself, see also Appendix \ref{app:A}. We find an extinction consistent with zero, suggesting that the SN is in front of the bright underlying \ion{H}{ii} region. Moreover, using emission line indicators useful for estimating the star-formation rate (SFR, \citealp{Kennicutt1994}) and the metallicity of the gas (i.e., the O3N2 and the N2 indices, \citealp{Marino2013}), we find a SFR of 0.08 M$_{\odot}$ per year and a metallicity of 12 + log(O/H) = 8.16 $\pm$ 0.18. This value is consistent with the those inferred for GRB-SN host galaxies, 12 + log(O/H) = 8.22$\pm$0.06 \citep{Japelj2018,Modjaz2019}.

\section{Discussion}\label{sec:4}

There is a striking analogy with the case of SN 2017iuk, where an early black-body component, characterized by high-velocity absorption lines in the optical spectra, was attributed to the presence of a cocoon component originating from the jet giving rise to GRB 171205A \citep{Izzo2019}. By comparing with SN 2017iuk, we can estimate the epoch of the collapse of the massive star associated with SN 2020bvc. Considering the observed value of $v_{\rm exp,1} \approx$ 70,000 km/s on Feb 5.54 UT and the evolution of the velocity in the cocoon of SN 2017iuk, we estimate that the core collapse would have happened $\approx 1.5 \pm 0.1$ days before the epoch of the FLOYDS spectrum, which results in the epoch of Feb 4.0 UT ($\pm$ 0.1 days) as the ''trigger'' time (Day 0) of the SN explosion. 

X-ray emission from SN 2020bvc was not detected by the \textit{Swift}-XRT in the first two days after the SN explosion: we report only an upper limit of $F_X < 3.23 \times 10^{-14}$ erg/cm$^2$/s at the position of the SN. An increase of the X-ray emission was detected $\approx$ three days after the discovery of the supernova. After the detection, the X-ray emission dropped following a power-law decay with a decay index $\alpha = 1.35 \pm 0.09$, which is consistent with the typical behavior observed in the late X-ray afterglows of GRBs \citep{Willingale2007}. The lack of a sufficient number of counts in the available X-ray observations prevents us from performing a detailed spectral analysis. However, we derive the hardness ratio, HR = $(H-S)/(H+S)$, where $S$ and $H$ are the number of counts in the (0.3 -- 2) keV and (2 -- 10) keV energy ranges, respectively (see Table \ref{tab:App3}). A weighted average of the HR values, gives HR = $-0.66\pm0.15$, indicating a soft spectrum consistent with a photon index  $1.9 < \gamma < 2.1$, where $N(\nu) \propto \nu^{-\gamma, } $\footnote{To show this, we simulated a series of XRT spectra varying the photon index $\gamma$ between 1.5 and 2.5 and the normalization to reproduce the observed X-ray fluxes, and assuming a Galactic and local absorption of N(H) = 10$^{20}$ cm$^{-2}$.}. The soft spectrum excludes a large amount of absorption, typical of interacting SNe
(e.g., like that seen in Type IIn SN2010jl: \citealt{2012ApJ...750L...2C}). 

\begin{figure}
\centering
\includegraphics[width=0.99\hsize,clip]{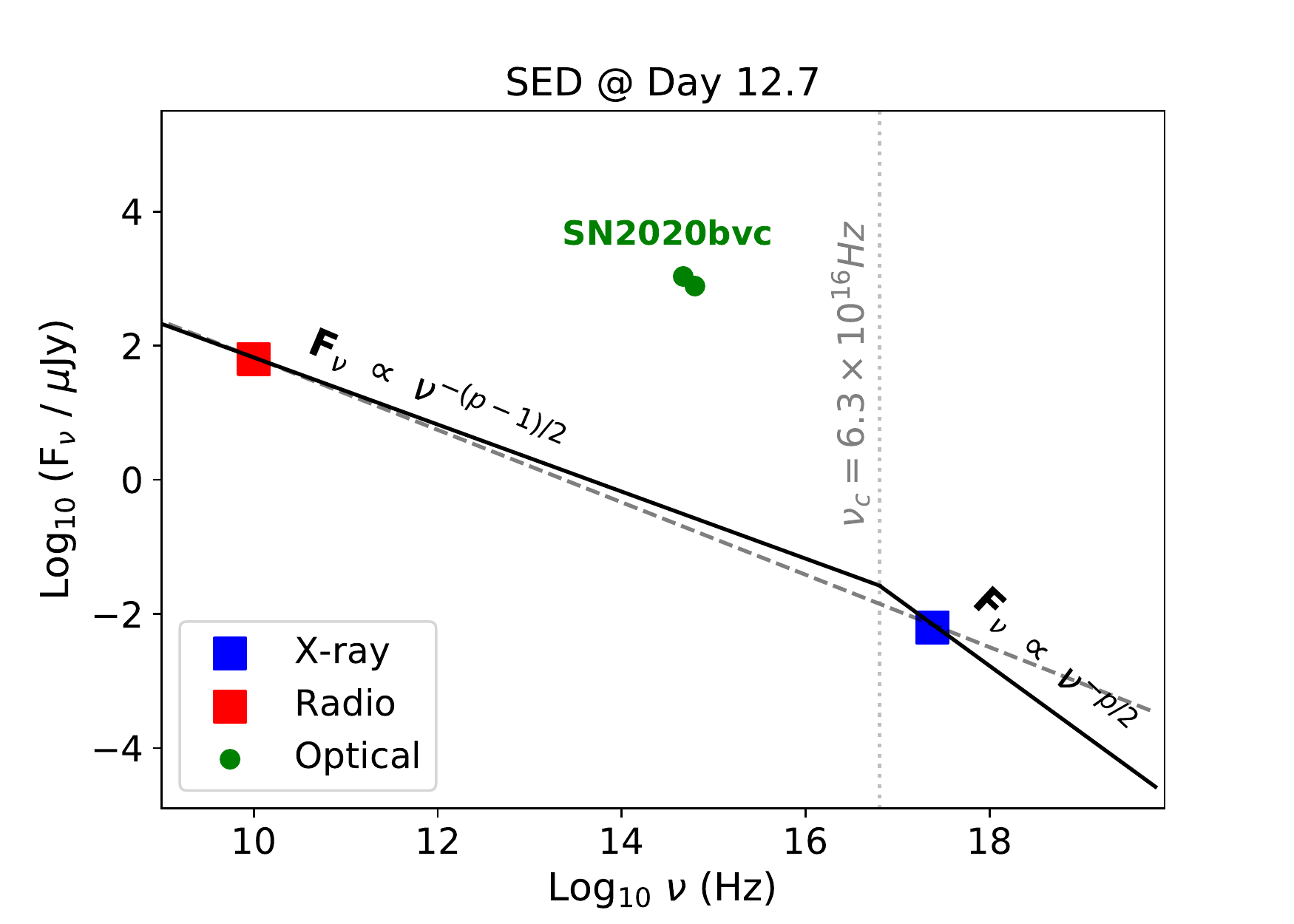}
\caption{
Multi-wavelength SED computed at Day 12.7 obtained using {\it CXO} data (blue data points -- this paper) and the VLA radio observations (red data points -- \citealp{Ho2020b}). We also included {\it ZTF} $g$ and $r$ data points, computed for the same epoch from interpolation of the {\it ZTF} light curve. X-rays and radio data are well explained within the slow cooling \citep{Sari1998,Granot2002} afterglow scenario (black line), with a cooling-break frequency at $\nu_c = 6.3\times10^{16}$ Hz. The dashed gray line shows the fit results using a single power-law segment with $p=2.08$. The optical data are well above the afterglow model due to the presence of the bright SN 2020bvc.
}
\label{fig:6}
\end{figure}

\begin{figure}
\centering
\includegraphics[width=0.99\hsize,clip]{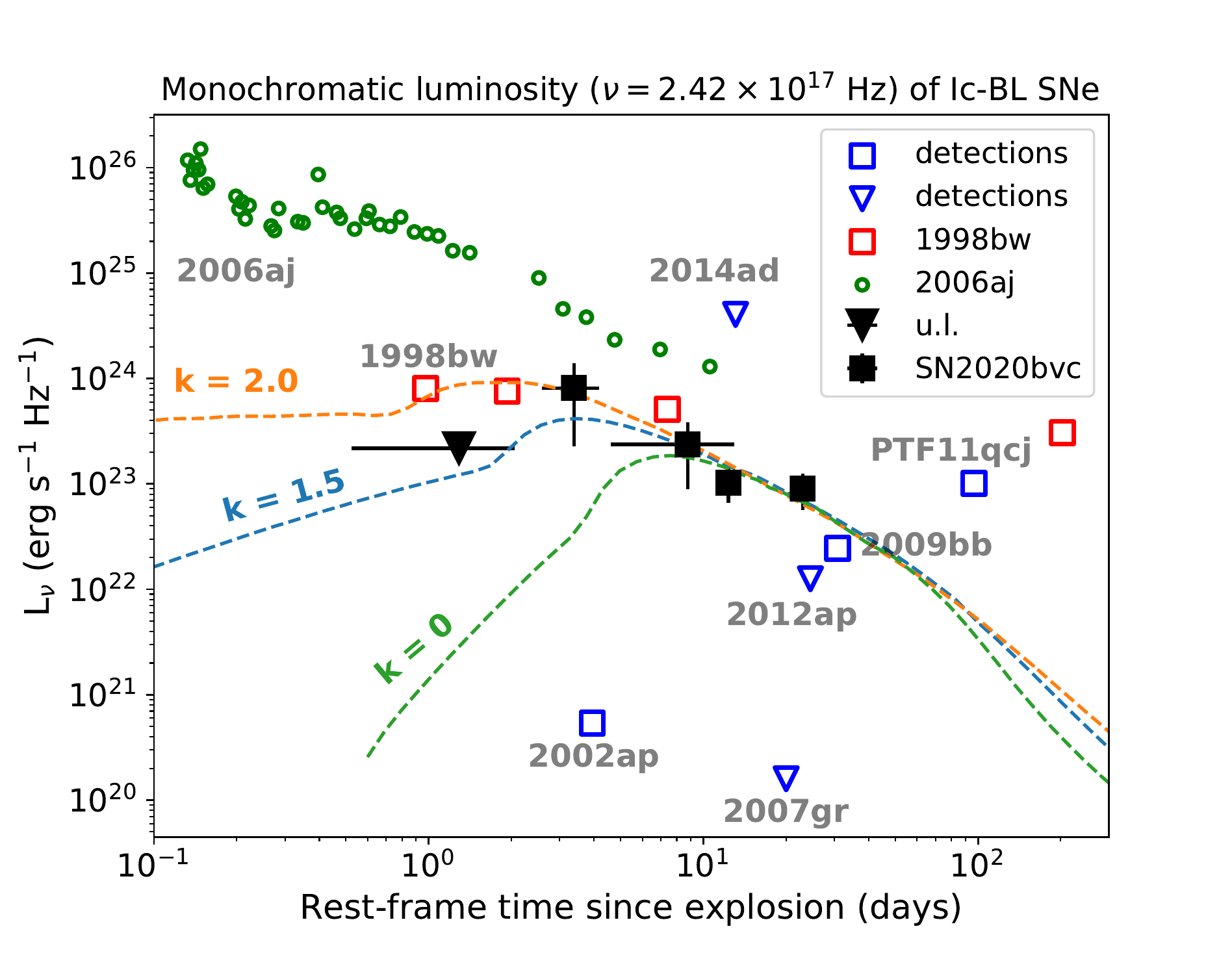}
\caption{
The monochromatic ($\nu = 2.42\times10^{17}$ Hz) luminosity evolution of SN 2020bvc in the first 16 days. The dashed lines represent simulations of an off-axis ($\theta = 23$ deg) X-ray afterglow characterized by a power-law circumburst density distribution, $\rho \propto R^{-k}$,
with index $k = 2.0$ (orange), $k = 1.5$ (blue) and $k = 0$ (green). The X-ray data and upper limits of SNe Ic-BL, including SN 1998bw associated with GRB 980425, reported with squares and triangles respectively, are from \citet{Margutti2014} and have been corrected for the (0.1--10 keV) energy range, assuming a power-law spectral model with photon index $\gamma = 2$. We also show the light curve of SN 2006aj associated with GRB 060218 \citep{Campana2006}.}
\label{fig:5}
\end{figure}

We can exclude that the X-rays are associated with emission from a choked jet: in this scenario, the cocoon is composed of the shocked stellar material that expands almost isotropically after breaking out from the progenitor star \citep{Nakar2017}. The observed emission peaks at $\sim$ 1 day, assuming an expansion velocity of $\sim$0.1$c$, and promptly fades similarly to the early cooling envelope emission reported for some SNe \citep{Nakar2015}. Instead, we have compared the total X-ray emission with simulations of an off-axis GRB afterglow light curve with the jet propagating in a stratified external medium \citep[][see also Figure \ref{fig:5}]{Granot2018}. We have found a reasonable match with the results obtained for a jet propagating at an off-axis angle of $\theta \approx 23$deg within an external density profile described by a power-law density distribution as a function of the distance from the GRB progenitor, $R^{-k}$, with the density power-law index $k = 1.5$, after re-scaling the theoretical curve for to distance of SN 2020bvc ($d = 120$ Mpc). We conclude that the X-ray emission is roughly consistent with the off-axis emission of a typical GRB with a jet energy $4\times 10^{50}$ erg, and microphysical parameters $\epsilon_e = \epsilon_B \sim 0.1$. 
The hardness ratio indicates that X-ray frequencies are in the $\nu^{-p/2}$  part of the spectrum. Using radio observations at $\sim$ 12.7 days \citep{Ho2020b}, for the case where $\nu_m < \nu_{radio}< \nu_c < \nu_{X-rays}$, we find that $\nu_c \sim 6.3 \times 10^{16}$ Hz for $p = 2$, see Fig.~\ref{fig:6}, which is consistent with the behavior observed for GRB afterglows at late times \citep[see, e.g.,][]{Kangas2020} and it is also consistent with the Radio SED presented in Fig.~5 in \citet{Ho2020b}. Alternatively, $ \nu_{radio} < \nu_m < \nu_c <\nu_{X-rays}$, in which case we cannot set strong constrains on the value of $p$, although the evolution of the slow-cooling synchrotron spectrum would suggest a delayed peak in the radio emission. This last scenario opens the possibility that the early radio emission is contaminated by another component, which could be the tail of the radio cocoon emission \citep{DeColle2018}. We also note that a good fit can be obtained with a less stratified environment and smaller observing angles (a $R^{-2}$ medium is not preferred by our models).

X-ray emission from the cocoon of a successful GRB is expected to peak a few hundred seconds after the explosion, then decaying quickly \citep{DeColle2018}, similarly to what seen in SN 2017iuk \citep{Izzo2019}, thus not representing an important contribution in the emission on the observed timescale ($\approx$ days).

In Figure \ref{fig:5} we compare the X-ray luminosity of SN 2020bvc with X-ray observations of low-luminosity GRB/SN, including SN 1998bw associated with GRB 980425 \citep{Galama1998} and SN 2006 associated with GRB 060218 \citep{Campana2006}, and other SNe Ic-BL. In the sample shown in Figure \ref{fig:5}, SN 2020bvc is one of the brightest SNe Ic-BL observed in X-rays, similar to the luminosity of SN 1998bw, but is $\sim$ one order of magnitude fainter than SN 2006aj. In this SN, the X-rays originate from a mildly-relativistic shock break-out whose luminosity and duration depend on the break-out radius $r_{SBO}$ \citep{Waxman2007,Nakar2015}. For typical values, $r_{SBO} \lesssim 10^{13}$ cm, we should have observed an X-ray emission few hours after the core collapse, then the lack of an X-ray detection in the first two days of the SN emission is not easily explained in the low-luminosity GRB scenario. We also note that SN 2002ap was observed at similar epochs than SN 2020bvc, but with a much lower luminosity ($\approx$ three orders of magnitude). This was attributed to emission from shocked circumstellar matter \citep{Soria2002}. We also note that the detection of SN 2009bb and PTF11qcj could be consistent with the simulations of an off-axis GRB afterglow \citep{DeColle2018b}. Radio observations of SN 2009bb have indeed demonstrated the presence of a bright radio afterglow, consistent with the emission from a relativistic outflow \citep{Soderberg2010}. 

The jet-cocoon emission and the X-ray afterglow suggest the presence of an off-axis collimated outflow in SN 2020bvc, which was responsible for the very early optical emission and for the observed behavior of the X-ray light curve. We have searched the \textit{Fermi} Gamma-ray Burst Monitor (GBM) and the \textit{INTEGRAL} archives to search for a possible gamma-ray counterpart detected by these instruments. At the expected time of the core collapse of SN 2020bvc (February 4.0 $\pm$ 0.1) we found only a sub-threshold \textit{Fermi}-GBM short burst event detected on February 3.8 UT at RA, Dec = (261.220, $-45.310$), which is $93.71$ degrees from the position of SN 2020bvc, much larger than the error radius of the GRB ($r = 6.91$ degrees). We have also checked the \textit{Swift} AFST log of observations\footnote{\url{https://www.swift.psu.edu/operations/obsSchedule.php}} at the time of the SN core collapse. \textit{Swift} was pointing at several targets in this time interval and the position of SN 2020bvc was partially covered from February 3.96 UT to February 4.08 UT, with a gap between February 4.02 and 4.04 UT. No high-energy triggers were reported by the Burst Alert Detector on-board \textit{Swift}. This is probably not surprising as the beaming angle for the gamma ray emission is expected to be far smaller.

The analysis of the emission lines in the late spectrum of SN 2020bvc suggests a low-metallicity environment at the SN location, which is thought to be a key ingredient for the Collapsar scenario \citep{Woosley2006}. This is because 
stellar evolution at low metallicities takes place with a significantly reduced mass-loss rate during the short lifetime of the pre-SN progenitor star \citep{Maeder2001} and implies faster rotating Fe cores \citep{Izzard2004}, which are expected to form jets in broad-lined core-collapse SN explosions.
We then conclude that SN 2020bvc represents the first orphan GRB afterglow detected through its associated SN emission and gives further credence to the idea that a wide variety of burst phenomenology from X-ray flashes \citep{Granot2005,Urata2015}
to a fraction of low-luminosity GRBs \citep{Ramirez-Ruiz2005}
could be attributable to a relatively standard type of event being viewed from different orientations. Upcoming wide-field optical and radio surveys have the potential to detect many more ($\sim$ 30--50 yr$^{-1}$, \citealt{Ghirlanda2015,Metzger2015}) off-axis GRB afterglows and their associated SNe, which will allow us to understand better the role of jets in SNe Ic-BL.

\begin{acknowledgements}
We thank the referee for his/her valuable comments, which definitely help to improve the quality of the paper. Based on observations collected at Copernico telescope (Asiago, Italy) of the INAF - Osservatorio Astronomico di Padova. We are grateful to P. Ochner and S. Benetti for allowing and executing the Asiago observations presented in this paper. This work was supported by a VILLUM FONDEN Investigator grant to JH (project number 16599) and by a VILLUM FONDEN Young Investigator grant to CG (project number 25501). KAA and ERR are supported by the Danish National Research Foundation (DNRF132). SIR gratefully acknowledges support from the Independent Research Fund Denmark via grant numbers DFF 4002-00275 and 8021-00130. FDC acknowledges support from UNAM-PAPIIT  grant
AG100820. Parts of this research were supported by the Australian Research Council Centre of Excellence for All Sky Astrophysics in 3 Dimensions (ASTRO 3D), through project number CE170100013.
\end{acknowledgements}

\bibliographystyle{aa}

\begin{thebibliography}{82}
\expandafter\ifx\csname natexlab\endcsname\relax\def\natexlab#1{#1}\fi

\bibitem[{{Arnaud}(1996)}]{Arnaud1996}
{Arnaud}, K.~A. 1996, in Astronomical Society of the Pacific Conference Series,
  Vol. 101, Astronomical Data Analysis Software and Systems V, ed. G.~H.
  {Jacoby} \& J.~{Barnes}, 17

\bibitem[{{Bellm} {et~al.}(2019){Bellm}, {Kulkarni}, {Graham}, {Dekany},
  {Smith}, {Riddle}, {Masci}, {Helou}, {Prince}, {Adams}, {Barbarino},
  {Barlow}, {Bauer}, {Beck}, {Belicki}, {Biswas}, {Blagorodnova}, {Bodewits},
  {Bolin}, {Brinnel}, {Brooke}, {Bue}, {Bulla}, {Burruss}, {Cenko}, {Chang},
  {Connolly}, {Coughlin}, {Cromer}, {Cunningham}, {De}, {Delacroix}, {Desai},
  {Duev}, {Eadie}, {Farnham}, {Feeney}, {Feindt}, {Flynn}, {Franckowiak},
  {Frederick}, {Fremling}, {Gal-Yam}, {Gezari}, {Giomi}, {Goldstein},
  {Golkhou}, {Goobar}, {Groom}, {Hacopians}, {Hale}, {Henning}, {Ho}, {Hover},
  {Howell}, {Hung}, {Huppenkothen}, {Imel}, {Ip}, {Ivezi{\'c}}, {Jackson},
  {Jones}, {Juric}, {Kasliwal}, {Kaspi}, {Kaye}, {Kelley}, {Kowalski},
  {Kramer}, {Kupfer}, {Landry}, {Laher}, {Lee}, {Lin}, {Lin}, {Lunnan},
  {Giomi}, {Mahabal}, {Mao}, {Miller}, {Monkewitz}, {Murphy}, {Ngeow},
  {Nordin}, {Nugent}, {Ofek}, {Patterson}, {Penprase}, {Porter}, {Rauch},
  {Rebbapragada}, {Reiley}, {Rigault}, {Rodriguez}, {van Roestel}, {Rusholme},
  {van Santen}, {Schulze}, {Shupe}, {Singer}, {Soumagnac}, {Stein}, {Surace},
  {Sollerman}, {Szkody}, {Taddia}, {Terek}, {Van Sistine}, {van Velzen},
  {Vestrand}, {Walters}, {Ward}, {Ye}, {Yu}, {Yan}, \& {Zolkower}}]{Bellm2019}
{Bellm}, E.~C., {Kulkarni}, S.~R., {Graham}, M.~J., {et~al.} 2019, \pasp, 131,
  018002

\bibitem[{{Breeveld} {et~al.}(2010){Breeveld}, {Curran}, {Hoversten}, {Koch},
  {Landsman}, {Marshall}, {Page}, {Poole}, {Roming}, {Smith}, {Still},
  {Yershov}, {Blustin}, {Brown}, {Gronwall}, {Holland}, {Kuin}, {McGowan},
  {Rosen}, {Boyd}, {Broos}, {Carter}, {Chester}, {Hancock}, {Huckle}, {Immler},
  {Ivanushkina}, {Kennedy}, {Mason}, {Morgan}, {Oates}, {de Pasquale},
  {Schady}, {Siegel}, \& {vand en Berk}}]{2010MNRAS.406.1687B}
{Breeveld}, A.~A., {Curran}, P.~A., {Hoversten}, E.~A., {et~al.} 2010, \mnras,
  406, 1687

\bibitem[{{Bromberg} {et~al.}(2011{\natexlab{a}}){Bromberg}, {Nakar}, \&
  {Piran}}]{Bromberg2011}
{Bromberg}, O., {Nakar}, E., \& {Piran}, T. 2011{\natexlab{a}}, \apjl, 739, L55

\bibitem[{{Bromberg} {et~al.}(2011{\natexlab{b}}){Bromberg}, {Nakar}, {Piran},
  \& {Sari}}]{Bromberg2011b}
{Bromberg}, O., {Nakar}, E., {Piran}, T., \& {Sari}, R. 2011{\natexlab{b}},
  \apj, 740, 100

\bibitem[{{Bromberg} {et~al.}(2012){Bromberg}, {Nakar}, {Piran}, \&
  {Sari}}]{Bromberg2012}
{Bromberg}, O., {Nakar}, E., {Piran}, T., \& {Sari}, R. 2012, \apj, 749, 110

\bibitem[{{Campana} {et~al.}(2006){Campana}, {Mangano}, {Blustin}, {Brown},
  {Burrows}, {Chincarini}, {Cummings}, {Cusumano}, {Della Valle}, {Malesani},
  {M{\'e}sz{\'a}ros}, {Nousek}, {Page}, {Sakamoto}, {Waxman}, {Zhang}, {Dai},
  {Gehrels}, {Immler}, {Marshall}, {Mason}, {Moretti}, {O'Brien}, {Osborne},
  {Page}, {Romano}, {Roming}, {Tagliaferri}, {Cominsky}, {Giommi}, {Godet},
  {Kennea}, {Krimm}, {Angelini}, {Barthelmy}, {Boyd}, {Palmer}, {Wells}, \&
  {White}}]{Campana2006}
{Campana}, S., {Mangano}, V., {Blustin}, A.~J., {et~al.} 2006, \nat, 442, 1008

\bibitem[{{Cano} {et~al.}(2017){Cano}, {Wang}, {Dai}, \& {Wu}}]{Cano2017}
{Cano}, Z., {Wang}, S.-Q., {Dai}, Z.-G., \& {Wu}, X.-F. 2017, Advances in
  Astronomy, 2017, 8929054

\bibitem[{{Cardelli} {et~al.}(1989){Cardelli}, {Clayton}, \&
  {Mathis}}]{Cardelli1989}
{Cardelli}, J.~A., {Clayton}, G.~C., \& {Mathis}, J.~S. 1989, \apj, 345, 245

\bibitem[{{Chakraborti} {et~al.}(2015){Chakraborti}, {Soderberg}, {Chomiuk},
  {Kamble}, {Yadav}, {Ray}, {Hurley}, {Margutti}, {Milisavljevic},
  {Bietenholz}, {Brunthaler}, {Pignata}, {Pian}, {Mazzali}, {Fransson},
  {Bartel}, {Hamuy}, {Levesque}, {MacFadyen}, {Dittmann}, {Krauss}, {Briggs},
  {Connaughton}, {Yamaoka}, {Takahashi}, {Ohno}, {Fukazawa}, {Tashiro},
  {Terada}, {Murakami}, {Goldsten}, {Barthelmy}, {Gehrels}, {Cummings},
  {Krimm}, {Palmer}, {Golenetskii}, {Aptekar}, {Frederiks}, {Svinkin}, {Cline},
  {Mitrofanov}, {Golovin}, {Litvak}, {Sanin}, {Boynton}, {Fellows}, {Harshman},
  {Enos}, {von Kienlin}, {Rau}, {Zhang}, \& {Savchenko}}]{Chakraborti2015}
{Chakraborti}, S., {Soderberg}, A., {Chomiuk}, L., {et~al.} 2015, \apj, 805,
  187

\bibitem[{{Chandra} {et~al.}(2012){Chandra}, {Chevalier}, {Irwin}, {Chugai},
  {Fransson}, \& {Soderberg}}]{2012ApJ...750L...2C}
{Chandra}, P., {Chevalier}, R.~A., {Irwin}, C.~M., {et~al.} 2012, \apjl, 750,
  L2

\bibitem[{{Corsi} {et~al.}(2016){Corsi}, {Gal-Yam}, {Kulkarni}, {Frail},
  {Mazzali}, {Cenko}, {Kasliwal}, {Cao}, {Horesh}, {Palliyaguru}, {Perley},
  {Laher}, {Taddia}, {Leloudas}, {Maguire}, {Nugent}, {Sollerman}, \&
  {Sullivan}}]{Corsi2016}
{Corsi}, A., {Gal-Yam}, A., {Kulkarni}, S.~R., {et~al.} 2016, \apj, 830, 42

\bibitem[{{De Colle} {et~al.}(2018{\natexlab{a}}){De Colle}, {Kumar}, \&
  {Aguilera-Dena}}]{DeColle2018}
{De Colle}, F., {Kumar}, P., \& {Aguilera-Dena}, D.~R. 2018{\natexlab{a}},
  \apj, 863, 32

\bibitem[{{De Colle} {et~al.}(2018{\natexlab{b}}){De Colle}, {Lu}, {Kumar},
  {Ramirez-Ruiz}, \& {Smoot}}]{DeColle2018b}
{De Colle}, F., {Lu}, W., {Kumar}, P., {Ramirez-Ruiz}, E., \& {Smoot}, G.
  2018{\natexlab{b}}, \mnras, 478, 4553

\bibitem[{{Fitzpatrick}(1999)}]{Fitzpatrick1999}
{Fitzpatrick}, E.~L. 1999, \pasp, 111, 63

\bibitem[{{Frail} {et~al.}(2001){Frail}, {Kulkarni}, {Sari}, {Djorgovski},
  {Bloom}, {Galama}, {Reichart}, {Berger}, {Harrison}, {Price}, {Yost},
  {Diercks}, {Goodrich}, \& {Chaffee}}]{Frail2001}
{Frail}, D.~A., {Kulkarni}, S.~R., {Sari}, R., {et~al.} 2001, \apjl, 562, L55

\bibitem[{{Gal-Yam}(2017)}]{Gal-Yam2017}
{Gal-Yam}, A. 2017, {Observational and Physical Classification of Supernovae},
  ed. A.~W. {Alsabti} \& P.~{Murdin}, 195

\bibitem[{{Galama} {et~al.}(1998){Galama}, {Vreeswijk}, {van Paradijs},
  {Kouveliotou}, {Augusteijn}, {B{\"o}hnhardt}, {Brewer}, {Doublier},
  {Gonzalez}, {Leibundgut}, {Lidman}, {Hainaut}, {Patat}, {Heise}, {in't Zand},
  {Hurley}, {Groot}, {Strom}, {Mazzali}, {Iwamoto}, {Nomoto}, {Umeda},
  {Nakamura}, {Young}, {Suzuki}, {Shigeyama}, {Koshut}, {Kippen}, {Robinson},
  {de Wildt}, {Wijers}, {Tanvir}, {Greiner}, {Pian}, {Palazzi}, {Frontera},
  {Masetti}, {Nicastro}, {Feroci}, {Costa}, {Piro}, {Peterson}, {Tinney},
  {Boyle}, {Cannon}, {Stathakis}, {Sadler}, {Begam}, \& {Ianna}}]{Galama1998}
{Galama}, T.~J., {Vreeswijk}, P.~M., {van Paradijs}, J., {et~al.} 1998, \nat,
  395, 670

\bibitem[{{Ghirlanda} {et~al.}(2013){Ghirlanda}, {Ghisellini}, {Salvaterra},
  {Nava}, {Burlon}, {Tagliaferri}, {Campana}, {D'Avanzo}, \&
  {Melandri}}]{Ghirlanda2013}
{Ghirlanda}, G., {Ghisellini}, G., {Salvaterra}, R., {et~al.} 2013, \mnras,
  428, 1410

\bibitem[{{Ghirlanda} {et~al.}(2015){Ghirlanda}, {Salvaterra}, {Campana},
  {Vergani}, {Japelj}, {Bernardini}, {Burlon}, {D'Avanzo}, {Melandri},
  {Gomboc}, {Nappo}, {Paladini}, {Pescalli}, {Salafia}, \&
  {Tagliaferri}}]{Ghirlanda2015}
{Ghirlanda}, G., {Salvaterra}, R., {Campana}, S., {et~al.} 2015, \aap, 578, A71

\bibitem[{{Graham} {et~al.}(2019){Graham}, {Kulkarni}, {Bellm}, {Adams},
  {Barbarino}, {Blagorodnova}, {Bodewits}, {Bolin}, {Brady}, {Cenko}, {Chang},
  {Coughlin}, {De}, {Eadie}, {Farnham}, {Feindt}, {Franckowiak}, {Fremling},
  {Gezari}, {Ghosh}, {Goldstein}, {Golkhou}, {Goobar}, {Ho}, {Huppenkothen},
  {Ivezi{\'c}}, {Jones}, {Juric}, {Kaplan}, {Kasliwal}, {Kelley}, {Kupfer},
  {Lee}, {Lin}, {Lunnan}, {Mahabal}, {Miller}, {Ngeow}, {Nugent}, {Ofek},
  {Prince}, {Rauch}, {van Roestel}, {Schulze}, {Singer}, {Sollerman}, {Taddia},
  {Yan}, {Ye}, {Yu}, {Barlow}, {Bauer}, {Beck}, {Belicki}, {Biswas}, {Brinnel},
  {Brooke}, {Bue}, {Bulla}, {Burruss}, {Connolly}, {Cromer}, {Cunningham},
  {Dekany}, {Delacroix}, {Desai}, {Duev}, {Feeney}, {Flynn}, {Frederick},
  {Gal-Yam}, {Giomi}, {Groom}, {Hacopians}, {Hale}, {Helou}, {Henning},
  {Hover}, {Hillenbrand}, {Howell}, {Hung}, {Imel}, {Ip}, {Jackson}, {Kaspi},
  {Kaye}, {Kowalski}, {Kramer}, {Kuhn}, {Landry}, {Laher}, {Mao}, {Masci},
  {Monkewitz}, {Murphy}, {Nordin}, {Patterson}, {Penprase}, {Porter},
  {Rebbapragada}, {Reiley}, {Riddle}, {Rigault}, {Rodriguez}, {Rusholme}, {van
  Santen}, {Shupe}, {Smith}, {Soumagnac}, {Stein}, {Surace}, {Szkody}, {Terek},
  {Van Sistine}, {van Velzen}, {Vestrand}, {Walters}, {Ward}, {Zhang}, \&
  {Zolkower}}]{Graham2019}
{Graham}, M.~J., {Kulkarni}, S.~R., {Bellm}, E.~C., {et~al.} 2019, \pasp, 131,
  078001

\bibitem[{{Granot} {et~al.}(2018){Granot}, {De Colle}, \&
  {Ramirez-Ruiz}}]{Granot2018}
{Granot}, J., {De Colle}, F., \& {Ramirez-Ruiz}, E. 2018, \mnras, 481, 2711

\bibitem[{{Granot} {et~al.}(2002){Granot}, {Panaitescu}, {Kumar}, \&
  {Woosley}}]{Granot2002}
{Granot}, J., {Panaitescu}, A., {Kumar}, P., \& {Woosley}, S.~E. 2002, \apjl,
  570, L61

\bibitem[{{Granot} {et~al.}(2005){Granot}, {Ramirez-Ruiz}, \&
  {Perna}}]{Granot2005}
{Granot}, J., {Ramirez-Ruiz}, E., \& {Perna}, R. 2005, \apj, 630, 1003

\bibitem[{{Guetta} \& {Della Valle}(2007)}]{Guetta2007}
{Guetta}, D. \& {Della Valle}, M. 2007, \apjl, 657, L73

\bibitem[{{HI4PI Collaboration} {et~al.}(2016){HI4PI Collaboration}, {Ben
  Bekhti}, {Fl{\"o}er}, {Keller}, {Kerp}, {Lenz}, {Winkel}, {Bailin},
  {Calabretta}, {Dedes}, {Ford}, {Gibson}, {Haud}, {Janowiecki}, {Kalberla},
  {Lockman}, {McClure-Griffiths}, {Murphy}, {Nakanishi}, {Pisano}, \&
  {Staveley-Smith}}]{2016A&A...594A.116H}
{HI4PI Collaboration}, {Ben Bekhti}, N., {Fl{\"o}er}, L., {et~al.} 2016, \aap,
  594, A116

\bibitem[{{Hiramatsu} {et~al.}(2020){Hiramatsu}, {Arcavi}, {Burke}, {Howell},
  {McCully}, {Pellegrino}, \& {Valenti}}]{Hiramatsu2020}
{Hiramatsu}, D., {Arcavi}, I., {Burke}, J., {et~al.} 2020, Transient Name
  Server Classification Report, 2020-403, 1

\bibitem[{{Hjorth} {et~al.}(2003){Hjorth}, {Sollerman}, {M{\o}ller}, {Fynbo},
  {Woosley}, {Kouveliotou}, {Tanvir}, {Greiner}, {Andersen}, {Castro-Tirado},
  {Castro Cer{\'o}n}, {Fruchter}, {Gorosabel}, {Jakobsson}, {Kaper}, {Klose},
  {Masetti}, {Pedersen}, {Pedersen}, {Pian}, {Palazzi}, {Rhoads}, {Rol}, {van
  den Heuvel}, {Vreeswijk}, {Watson}, \& {Wijers}}]{Hjorth2003}
{Hjorth}, J., {Sollerman}, J., {M{\o}ller}, P., {et~al.} 2003, \nat, 423, 847

\bibitem[{{Ho} {et~al.}(2020{\natexlab{a}}){Ho}, {Cenko}, {Perley}, {Corsi}, \&
  {Brightman}}]{Ho2020a}
{Ho}, A.~Y.~Q., {Cenko}, B., {Perley}, D., {Corsi}, A., \& {Brightman}, M.
  2020{\natexlab{a}}, Transient Name Server AstroNote, 45, 1

\bibitem[{{Ho} {et~al.}(2019){Ho}, {Corsi}, {Cenko}, {Taddia}, {Kulkarni},
  {Adams}, {De}, {Dekany}, {Frederiks}, {Fremling}, {Golkhou}, {Kupfer},
  {Laher}, {Mahabal}, {Masci}, {Miller}, {Neill}, {Reiley}, {Riddle},
  {Ridnaia}, {Rusholme}, {Sharma}, {Sollerman}, {Soumagnac}, {Svinkin}, \&
  {Shupe}}]{Ho2019}
{Ho}, A. Y.~Q., {Corsi}, A., {Cenko}, S.~B., {et~al.} 2019, arXiv e-prints,
  arXiv:1912.10354

\bibitem[{{Ho} {et~al.}(2020{\natexlab{b}}){Ho}, {Kulkarni}, {Perley}, {Cenko},
  {Corsi}, {Schulze}, {Lunnan}, {Sollerman}, {Gal-Yam}, {Anand}, {Barbarino},
  {Bellm}, {Bruch}, {Burns}, {De}, {Dekany}, {Delacroix}, {Duev}, {Fremling},
  {Goldstein}, {Golkhou}, {Graham}, {Hale}, {Kasliwal}, {Kupfer}, {Laher},
  {Martikainen}, {Masci}, {Neill}, {Rusholme}, {Shupe}, {Soumagnac},
  {Strotjohann}, {Taggart}, {Tartaglia}, {Yan}, \& {Zolkower}}]{Ho2020b}
{Ho}, A.~Y.~Q., {Kulkarni}, S.~R., {Perley}, D.~A., {et~al.}
  2020{\natexlab{b}}, arXiv e-prints, arXiv:2004.10406

\bibitem[{{Irwin} {et~al.}(2019){Irwin}, {Nakar}, \& {Piran}}]{Irwin2019}
{Irwin}, C.~M., {Nakar}, E., \& {Piran}, T. 2019, \mnras, 489, 2844

\bibitem[{{Izzard} {et~al.}(2004){Izzard}, {Ramirez-Ruiz}, \&
  {Tout}}]{Izzard2004}
{Izzard}, R.~G., {Ramirez-Ruiz}, E., \& {Tout}, C.~A. 2004, \mnras, 348, 1215

\bibitem[{{Izzo} {et~al.}(2019){Izzo}, {de Ugarte Postigo}, {Maeda},
  {Th{\"o}ne}, {Kann}, {Della Valle}, {Sagues Carracedo}, {Micha{\l}owski},
  {Schady}, {Schmidl}, {Selsing}, {Starling}, {Suzuki}, {Bensch}, {Bolmer},
  {Campana}, {Cano}, {Covino}, {Fynbo}, {Hartmann}, {Heintz}, {Hjorth},
  {Japelj}, {Kami{\'n}ski}, {Kaper}, {Kouveliotou}, {Kru{\.Z}y{\'n}ski},
  {Kwiatkowski}, {Leloudas}, {Levan}, {Malesani}, {Micha{\l}owski},
  {Piranomonte}, {Pugliese}, {Rossi}, {S{\'a}nchez-Ram{\'\i}rez}, {Schulze},
  {Steeghs}, {Tanvir}, {Ulaczyk}, {Vergani}, \& {Wiersema}}]{Izzo2019}
{Izzo}, L., {de Ugarte Postigo}, A., {Maeda}, K., {et~al.} 2019, \nat, 565, 324

\bibitem[{{Japelj} {et~al.}(2018){Japelj}, {Vergani}, {Salvaterra}, {Renzo},
  {Zapartas}, {de Mink}, {Kaper}, \& {Zibetti}}]{Japelj2018}
{Japelj}, J., {Vergani}, S.~D., {Salvaterra}, R., {et~al.} 2018, \aap, 617,
  A105

\bibitem[{{Kaneko} {et~al.}(2007){Kaneko}, {Ramirez-Ruiz}, {Granot},
  {Kouveliotou}, {Woosley}, {Patel}, {Rol}, {in 't Zand}, {van der Horst},
  {Wijers}, \& {Strom}}]{Kaneko2007}
{Kaneko}, Y., {Ramirez-Ruiz}, E., {Granot}, J., {et~al.} 2007, \apj, 654, 385

\bibitem[{{Kangas} {et~al.}(2020){Kangas}, {Fruchter}, {Cenko}, {Corsi},
  {Postigo}, {Pe'er}, {Vogel}, {Cucchiara}, {Gompertz}, {Graham}, {Levan},
  {Misra}, {Perley}, {Racusin}, \& {Tanvir}}]{Kangas2020}
{Kangas}, T., {Fruchter}, A.~S., {Cenko}, S.~B., {et~al.} 2020, \apj, 894, 43

\bibitem[{{Kennicutt} {et~al.}(1994){Kennicutt}, {Tamblyn}, \&
  {Congdon}}]{Kennicutt1994}
{Kennicutt}, Robert~C., J., {Tamblyn}, P., \& {Congdon}, C.~E. 1994, \apj, 435,
  22

\bibitem[{{Komissarov} {et~al.}(2010){Komissarov}, {Vlahakis}, \&
  {K{\"o}nigl}}]{Komissarov2010}
{Komissarov}, S.~S., {Vlahakis}, N., \& {K{\"o}nigl}, A. 2010, \mnras, 407, 17

\bibitem[{{Kumar} \& {Granot}(2003)}]{Kumar2003}
{Kumar}, P. \& {Granot}, J. 2003, \apj, 591, 1075

\bibitem[{{Kumar} \& {Zhang}(2015)}]{Kumar2015}
{Kumar}, P. \& {Zhang}, B. 2015, \physrep, 561, 1

\bibitem[{{Lazzati} {et~al.}(2012){Lazzati}, {Morsony}, {Blackwell}, \&
  {Begelman}}]{Lazzati2012}
{Lazzati}, D., {Morsony}, B.~J., {Blackwell}, C.~H., \& {Begelman}, M.~C. 2012,
  \apj, 750, 68

\bibitem[{{Levan} {et~al.}(2005){Levan}, {Nugent}, {Fruchter}, {Burud},
  {Branch}, {Rhoads}, {Castro-Tirado}, {Gorosabel}, {Castro Cer{\'o}n},
  {Thorsett}, {Kouveliotou}, {Golenetskii}, {Fynbo}, {Garnavich}, {Holland},
  {Hjorth}, {M{\o}ller}, {Pian}, {Tanvir}, {Ulanov}, {Wijers}, \&
  {Woosley}}]{Levan2005}
{Levan}, A., {Nugent}, P., {Fruchter}, A., {et~al.} 2005, \apj, 624, 880

\bibitem[{{MacFadyen} {et~al.}(2001){MacFadyen}, {Woosley}, \&
  {Heger}}]{MacFadyen2001}
{MacFadyen}, A.~I., {Woosley}, S.~E., \& {Heger}, A. 2001, \apj, 550, 410

\bibitem[{{Maeder} \& {Meynet}(2001)}]{Maeder2001}
{Maeder}, A. \& {Meynet}, G. 2001, \aap, 373, 555

\bibitem[{{Marcote} {et~al.}(2019){Marcote}, {Nimmo}, {Salafia}, {Paragi},
  {Hessels}, {Petroff}, \& {Karuppusamy}}]{Marcote2019}
{Marcote}, B., {Nimmo}, K., {Salafia}, O.~S., {et~al.} 2019, \apjl, 876, L14

\bibitem[{{Margutti} {et~al.}(2014){Margutti}, {Milisavljevic}, {Soderberg},
  {Guidorzi}, {Morsony}, {Sanders}, {Chakraborti}, {Ray}, {Kamble}, {Drout},
  {Parrent}, {Zauderer}, \& {Chomiuk}}]{Margutti2014}
{Margutti}, R., {Milisavljevic}, D., {Soderberg}, A.~M., {et~al.} 2014, \apj,
  797, 107

\bibitem[{{Marino} {et~al.}(2013){Marino}, {Rosales-Ortega}, {S{\'a}nchez},
  {Gil de Paz}, {V{\'{\i}}lchez}, {Miralles-Caballero}, {Kehrig},
  {P{\'e}rez-Montero}, {Stanishev}, {Iglesias-P{\'a}ramo}, {D{\'{\i}}az},
  {Castillo-Morales}, {Kennicutt}, {L{\'o}pez-S{\'a}nchez}, {Galbany},
  {Garc{\'{\i}}a-Benito}, {Mast}, {Mendez-Abreu}, {Monreal-Ibero}, {Husemann},
  {Walcher}, {Garc{\'{\i}}a-Lorenzo}, {Masegosa}, {Del Olmo Orozco},
  {Mour{\~a}o}, {Ziegler}, {Moll{\'a}}, {Papaderos},
  {S{\'a}nchez-Bl{\'a}zquez}, {Gonz{\'a}lez Delgado}, {Falc{\'o}n-Barroso},
  {Roth}, {van de Ven}, \& {Califa Team}}]{Marino2013}
{Marino}, R.~A., {Rosales-Ortega}, F.~F., {S{\'a}nchez}, S.~F., {et~al.} 2013,
  A\&A, 559, A114

\bibitem[{{Marongiu} {et~al.}(2019){Marongiu}, {Guidorzi}, {Margutti},
  {Coppejans}, {Martone}, \& {Kamble}}]{Marongiu2019}
{Marongiu}, M., {Guidorzi}, C., {Margutti}, R., {et~al.} 2019, \apj, 879, 89

\bibitem[{{M{\'e}sz{\'a}ros} \& {Rees}(2001)}]{Meszaros2001}
{M{\'e}sz{\'a}ros}, P. \& {Rees}, M.~J. 2001, \apjl, 556, L37

\bibitem[{{Metzger} {et~al.}(2015){Metzger}, {Williams}, \&
  {Berger}}]{Metzger2015}
{Metzger}, B.~D., {Williams}, P.~K.~G., \& {Berger}, E. 2015, \apj, 806, 224

\bibitem[{{Milisavljevic} {et~al.}(2015){Milisavljevic}, {Margutti}, {Parrent},
  {Soderberg}, {Fesen}, {Mazzali}, {Maeda}, {Sanders}, {Cenko}, {Silverman},
  {Filippenko}, {Kamble}, {Chakraborti}, {Drout}, {Kirshner}, {Pickering},
  {Kawabata}, {Hattori}, {Hsiao}, {Stritzinger}, {Marion}, {Vinko}, \&
  {Wheeler}}]{Milisavljevic2015}
{Milisavljevic}, D., {Margutti}, R., {Parrent}, J.~T., {et~al.} 2015, \apj,
  799, 51

\bibitem[{{Modjaz} {et~al.}(2019){Modjaz}, {Bianco}, {Siwek}, {Huang},
  {Perley}, {Fierroz}, {Liu}, {Arcavi}, {Gal-Yam}, {Blagorodnova}, {Cenko},
  {Filippenko}, {Kasliwal}, {Kulkarni}, {Schulze}, {Taggart}, \&
  {Zhen}}]{Modjaz2019}
{Modjaz}, M., {Bianco}, F.~B., {Siwek}, M., {et~al.} 2019, arXiv e-prints,
  arXiv:1901.00872

\bibitem[{{Modjaz} {et~al.}(2016){Modjaz}, {Liu}, {Bianco}, \&
  {Graur}}]{Modjaz2016}
{Modjaz}, M., {Liu}, Y.~Q., {Bianco}, F.~B., \& {Graur}, O. 2016, \apj, 832,
  108

\bibitem[{{Moretti} {et~al.}(2004){Moretti}, {Campana}, {Tagliaferri}, {Abbey},
  {Ambrosi}, {Angelini}, {Beardmore}, {Br{\"a}uninger}, {Burkert}, {Burrows},
  {Capalbi}, {Chincarini}, {Citterio}, {Cusumano}, {Freyberg}, {Giommi},
  {Hartner}, {Hill}, {Mori}, {Morris}, {Mukerjee}, {Nousek}, {Osborne},
  {Short}, {Tamburelli}, {Watson}, \& {Wells}}]{2004SPIE.5165..232M}
{Moretti}, A., {Campana}, S., {Tagliaferri}, G., {et~al.} 2004, Society of
  Photo-Optical Instrumentation Engineers (SPIE) Conference Series, Vol. 5165,
  {SWIFT XRT point spread function measured at the Panter end-to-end tests},
  ed. K.~A. {Flanagan} \& O.~H.~W. {Siegmund}, 232--240

\bibitem[{{Nakar}(2015)}]{Nakar2015}
{Nakar}, E. 2015, \apj, 807, 172

\bibitem[{{Nakar} \& {Piran}(2017)}]{Nakar2017}
{Nakar}, E. \& {Piran}, T. 2017, \apj, 834, 28

\bibitem[{{Osterbrock} \& {Ferland}(2006)}]{Osterbrock2006}
{Osterbrock}, D.~E. \& {Ferland}, G.~J. 2006, {Astrophysics of gaseous nebulae
  and active galactic nuclei}

\bibitem[{{Perley} {et~al.}(2020){Perley}, {Schulze}, \& {Bruch}}]{Perley2020}
{Perley}, D., {Schulze}, S., \& {Bruch}, R. 2020, Transient Name Server
  AstroNote, 37, 1

\bibitem[{Piran(2005)}]{Piran2005}
Piran, T. 2005, Rev. Mod. Phys., 76, 1143

\bibitem[{{Ramirez-Ruiz} {et~al.}(2002){Ramirez-Ruiz}, {Celotti}, \&
  {Rees}}]{Ramirez-Ruiz2002}
{Ramirez-Ruiz}, E., {Celotti}, A., \& {Rees}, M.~J. 2002, \mnras, 337, 1349

\bibitem[{{Ramirez-Ruiz} {et~al.}(2005){Ramirez-Ruiz}, {Granot}, {Kouveliotou},
  {Woosley}, {Patel}, \& {Mazzali}}]{Ramirez-Ruiz2005}
{Ramirez-Ruiz}, E., {Granot}, J., {Kouveliotou}, C., {et~al.} 2005, \apjl, 625,
  L91

\bibitem[{{Rhoads}(1997)}]{Rhoads1997}
{Rhoads}, J.~E. 1997, \apjl, 487, L1

\bibitem[{{Sahu} {et~al.}(2018){Sahu}, {Anupama}, {Chakradhari}, {Srivastav},
  {Tanaka}, {Maeda}, \& {Nomoto}}]{Sahu2018}
{Sahu}, D.~K., {Anupama}, G.~C., {Chakradhari}, N.~K., {et~al.} 2018, \mnras,
  475, 2591

\bibitem[{{Sanders} {et~al.}(2012){Sanders}, {Soderberg}, {Levesque}, {Foley},
  {Chornock}, {Milisavljevic}, {Margutti}, {Berger}, {Drout}, {Czekala}, \&
  {Dittmann}}]{Sanders2012}
{Sanders}, N.~E., {Soderberg}, A.~M., {Levesque}, E.~M., {et~al.} 2012, \apj,
  758, 132

\bibitem[{{Sari} {et~al.}(1998){Sari}, {Piran}, \& {Narayan}}]{Sari1998}
{Sari}, R., {Piran}, T., \& {Narayan}, R. 1998, \apjl, 497, L17

\bibitem[{{Schlafly} \& {Finkbeiner}(2011)}]{Schlafly2011}
{Schlafly}, E.~F. \& {Finkbeiner}, D.~P. 2011, \apj, 737, 103

\bibitem[{{Shappee} {et~al.}(2014){Shappee}, {Prieto}, {Grupe}, {Kochanek},
  {Stanek}, {De Rosa}, {Mathur}, {Zu}, {Peterson}, {Pogge}, {Komossa}, {Im},
  {Jencson}, {Holoien}, {Basu}, {Beacom}, {Szczygie{\l}}, {Brimacombe},
  {Adams}, {Campillay}, {Choi}, {Contreras}, {Dietrich}, {Dubberley},
  {Elphick}, {Foale}, {Giustini}, {Gonzalez}, {Hawkins}, {Howell}, {Hsiao},
  {Koss}, {Leighly}, {Morrell}, {Mudd}, {Mullins}, {Nugent}, {Parrent},
  {Phillips}, {Pojmanski}, {Rosing}, {Ross}, {Sand}, {Terndrup}, {Valenti},
  {Walker}, \& {Yoon}}]{2014ApJ...788...48S}
{Shappee}, B.~J., {Prieto}, J.~L., {Grupe}, D., {et~al.} 2014, \apj, 788, 48

\bibitem[{{Soderberg} {et~al.}(2010){Soderberg}, {Chakraborti}, {Pignata},
  {Chevalier}, {Chandra}, {Ray}, {Wieringa}, {Copete}, {Chaplin},
  {Connaughton}, {Barthelmy}, {Bietenholz}, {Chugai}, {Stritzinger}, {Hamuy},
  {Fransson}, {Fox}, {Levesque}, {Grindlay}, {Challis}, {Foley}, {Kirshner},
  {Milne}, \& {Torres}}]{Soderberg2010}
{Soderberg}, A.~M., {Chakraborti}, S., {Pignata}, G., {et~al.} 2010, \nat, 463,
  513

\bibitem[{{Soderberg} {et~al.}(2006){Soderberg}, {Nakar}, {Berger}, \&
  {Kulkarni}}]{Soderberg2006}
{Soderberg}, A.~M., {Nakar}, E., {Berger}, E., \& {Kulkarni}, S.~R. 2006, \apj,
  638, 930

\bibitem[{{Sollerman} {et~al.}(2006){Sollerman}, {Jaunsen}, {Fynbo}, {Hjorth},
  {Jakobsson}, {Stritzinger}, {F{\'e}ron}, {Laursen}, {Ovaldsen}, {Selj},
  {Th{\"o}ne}, {Xu}, {Davis}, {Gorosabel}, {Watson}, {Duro}, {Ilyin}, {Jensen},
  {Lysfjord}, {Marquart}, {Nielsen}, {N{\"a}r{\"a}nen}, {Schwarz}, {Walch},
  {Wold}, \& {{\"O}stlin}}]{Sollerman2006}
{Sollerman}, J., {Jaunsen}, A.~O., {Fynbo}, J.~P.~U., {et~al.} 2006, \aap, 454,
  503

\bibitem[{{Soria} \& {Kong}(2002)}]{Soria2002}
{Soria}, R. \& {Kong}, A. K.~H. 2002, \apjl, 572, L33

\bibitem[{{Stanek}(2020)}]{Stanek2020}
{Stanek}, K.~Z. 2020, Transient Name Server Discovery Report, 2020-381, 1

\bibitem[{{Stanek} {et~al.}(2003){Stanek}, {Matheson}, {Garnavich}, {Martini},
  {Berlind}, {Caldwell}, {Challis}, {Brown}, {Schild}, {Krisciunas}, {Calkins},
  {Lee}, {Hathi}, {Jansen}, {Windhorst}, {Echevarria}, {Eisenstein}, {Pindor},
  {Olszewski}, {Harding}, {Holland }, \& {Bersier}}]{2003ApJ...591L..17S}
{Stanek}, K.~Z., {Matheson}, T., {Garnavich}, P.~M., {et~al.} 2003, \apjl, 591,
  L17

\bibitem[{{Stevance} {et~al.}(2017){Stevance}, {Maund}, {Baade}, {H{\"o}flich},
  {Howerton}, {Patat}, {Rose}, {Spyromilio}, {Wheeler}, \&
  {Wang}}]{Stevance2017}
{Stevance}, H.~F., {Maund}, J.~R., {Baade}, D., {et~al.} 2017, \mnras, 469,
  1897

\bibitem[{{Taddia} {et~al.}(2019){Taddia}, {Sollerman}, {Fremling},
  {Barbarino}, {Karamehmetoglu}, {Arcavi}, {Cenko}, {Filippenko}, {Gal-Yam},
  {Hiramatsu}, {Hosseinzadeh}, {Howell}, {Kulkarni}, {Laher}, {Lunnan},
  {Masci}, {Nugent}, {Nyholm}, {Perley}, {Quimby}, \& {Silverman}}]{Taddia2019}
{Taddia}, F., {Sollerman}, J., {Fremling}, C., {et~al.} 2019, \aap, 621, A71

\bibitem[{{Tchekhovskoy} {et~al.}(2009){Tchekhovskoy}, {McKinney}, \&
  {Narayan}}]{Tchekhovskoy2009}
{Tchekhovskoy}, A., {McKinney}, J.~C., \& {Narayan}, R. 2009, \apj, 699, 1789

\bibitem[{{Tody}(1986)}]{Tody1986}
{Tody}, D. 1986, Society of Photo-Optical Instrumentation Engineers (SPIE)
  Conference Series, Vol. 627, {The IRAF Data Reduction and Analysis System},
  ed. D.~L. {Crawford}, 733

\bibitem[{{Urata} {et~al.}(2015){Urata}, {Huang}, {Yamazaki}, \&
  {Sakamoto}}]{Urata2015}
{Urata}, Y., {Huang}, K., {Yamazaki}, R., \& {Sakamoto}, T. 2015, \apj, 806,
  222

\bibitem[{{Waxman} {et~al.}(2007){Waxman}, {M{\'e}sz{\'a}ros}, \&
  {Campana}}]{Waxman2007}
{Waxman}, E., {M{\'e}sz{\'a}ros}, P., \& {Campana}, S. 2007, \apj, 667, 351

\bibitem[{{Willingale} {et~al.}(2007){Willingale}, {O'Brien}, {Osborne},
  {Godet}, {Page}, {Goad}, {Burrows}, {Zhang}, {Rol}, {Gehrels}, \&
  {Chincarini}}]{Willingale2007}
{Willingale}, R., {O'Brien}, P.~T., {Osborne}, J.~P., {et~al.} 2007, \apj, 662,
  1093

\bibitem[{{Woosley} \& {Bloom}(2006)}]{Woosley2006}
{Woosley}, S.~E. \& {Bloom}, J.~S. 2006, \araa, 44, 507

\end{thebibliography}

\begin{appendix}

\section*{A. The local environment of SN 2020bvc}\label{app:A}

We investigated the physical properties of the gas surrounding the location of SN 2020bvc from an analysis of the nebular emission lines visible in the spectrum. We clearly detect Balmer lines (H$\alpha$, H$\beta$, H$\gamma$), the [\ion{O}{iii}] 4959,5007 $\AA$ lines, and a very faint [\ion{N}{ii}] 6584 $\AA$ line. We report the fluxes of these lines in Table \ref{tab:1}. From an analysis of the Balmer lines we have estimated the local extinction using the Balmer decrement ratio. The observed ratio $(H\alpha / H\beta)_{\rm obs} = 2.88\pm0.22$ is consistent with the theoretical value of $(H\alpha / H\beta)_{\rm th} = 2.86$, which is obtained from a typical \ion{H}{ii} temperature of $T =$ 10,000 K and an electron density $n_e = 10^2$ cm$^{-3}$ for Case B recombination \citep{Osterbrock2006}. This suggests that the SN is located in front of the bright \ion{H}{ii} region in UGC 9379 and, in particular, it suggests a local extinction consistent with zero, $E(B-V)_{\rm int} = 0.01\pm0.01$ mag. From the luminosity of the H$\alpha$ line we have estimated the SFR using the \citet{Kennicutt1994} formulation: from an inferred luminosity of $L = 1.08 \times 10^{40}$ erg/s, we obtain a SFR of 0.08 M$_{\odot}$ per year. Finally, we have estimated the gas metallicity using the emission line ratio indicators N2 and O3N2 in the formulation provided by \citet{Marino2013}. Using the values reported in Table \ref{tab:1}, we obtain 12 + log(O/H) = 8.16 $\pm$ 0.18 using the O3N2 indicator and 12 + log(O/H) = 8.36 $\pm$ 0.16 using the N2 indicator. The values obtained from the N2 and O3N2 indicators are consistent within their uncertainties. If we use the O3N2 value, the metallicity inferred for the SN 2020bvc environment is consistent with the value inferred for GRB-SNe host galaxies, 12 + log(O/H) = 8.22$\pm$0.06 \citep{Japelj2018,Modjaz2019}.

\begin{table}
\centering
\caption{Observed fluxes for the nebular emission lines measured in the Asiago spectrum of SN 2020bvc.}
\begin{tabular}{lcc}
\hline\hline
Line & Flux  & EW\\
 & (10$^{-15}$ erg/cm$^2$/s) & ($\AA$) \\
\hline
H$\gamma$  & 1.20$\pm$0.19 & 4.42$\pm$1.58\\
H$\beta$  & 2.42$\pm$0.38 & 10.9$\pm$2.48 \\
$[\ion{O}{iii}]$ 5007  & 7.28$\pm$1.14 & 23.64$\pm$3.64 \\
H$\alpha$ & 6.98$\pm$1.09 & 18.9 $\pm$ 3.2 \\
$[\ion{N}{ii}]$ 6584 & 0.37$\pm$0.05 & 0.58$\pm$0.37 \\
\hline												
\end{tabular}
\label{tab:1}
\end{table}

\section*{B. SED analysis}\label{app:2}

\begin{figure*}[h!]
\centering
\includegraphics[width=0.35\hsize,angle=270,clip]{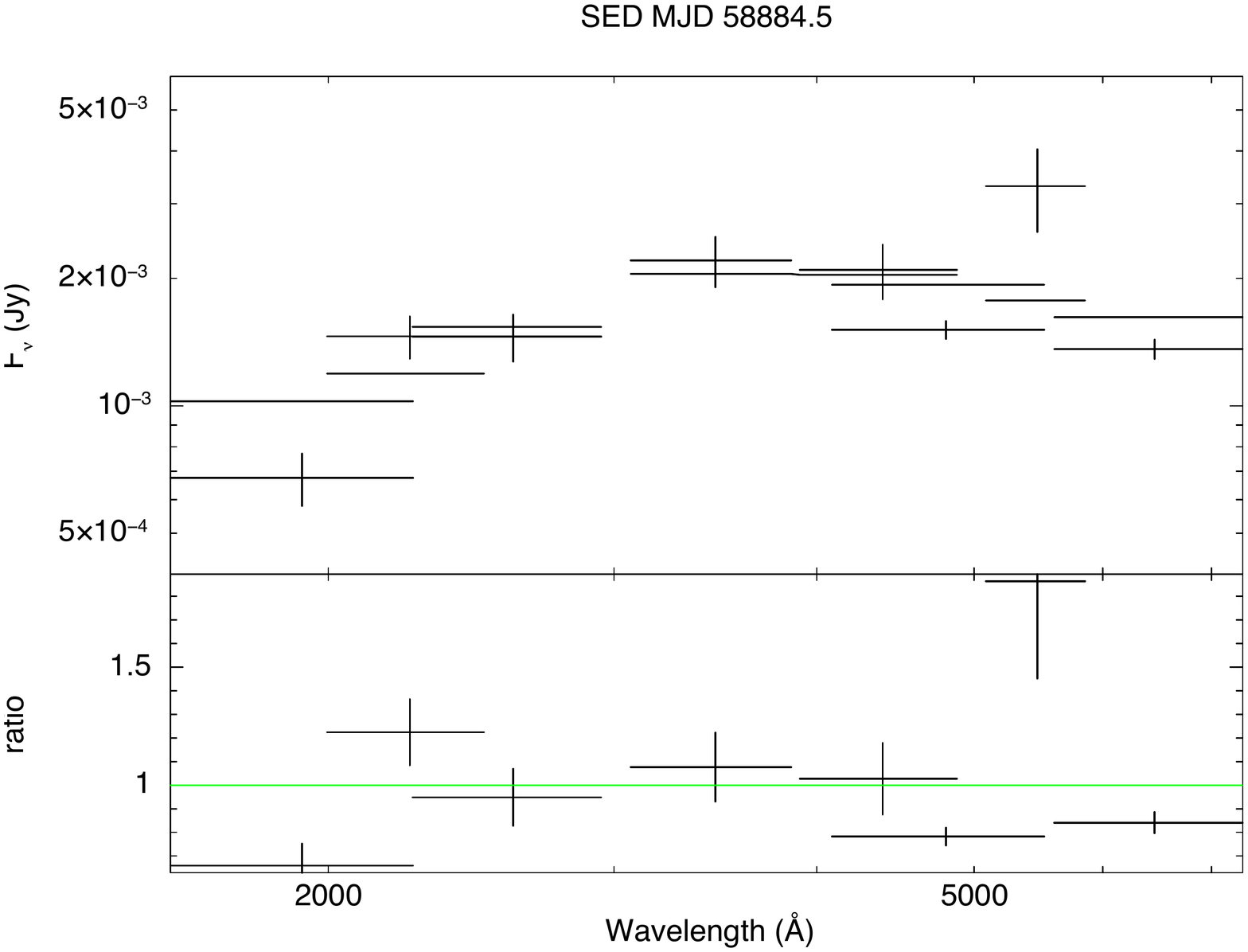}
\includegraphics[width=0.35\hsize,angle=270,clip]{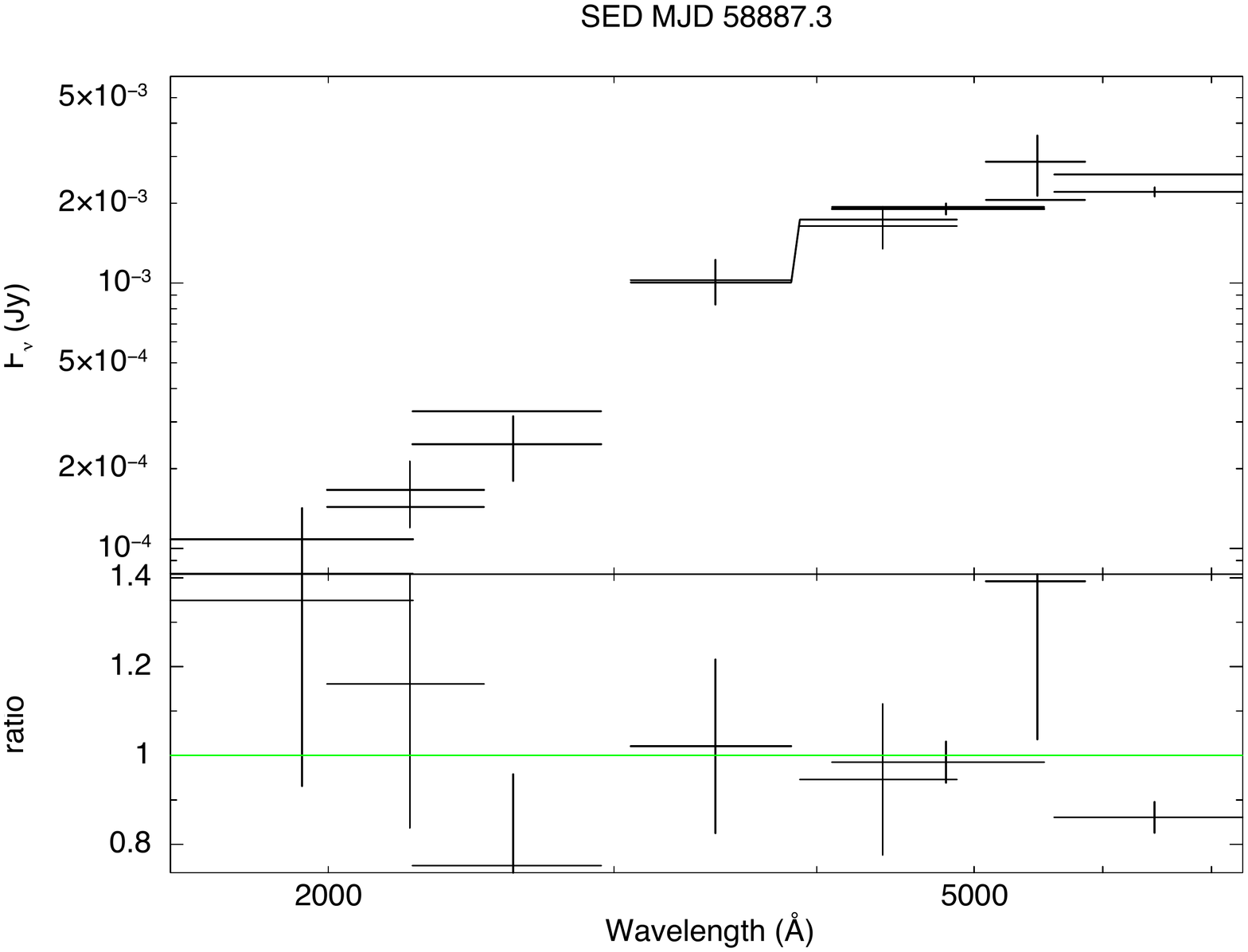}
\caption{The {\it Swift} and {\it ZTF} data of SN 2020bvc obtained on MJD = 58884.0 ({\it left panel}) and MJD = 58887.3  ({\it right panel}) with the best-fit results obtained using an absorbed black body model (see text).}
\label{fig:App1}
\end{figure*}

We have used the {\it Swift}-UVOT observations to build two SEDs at the earliest epochs in order to constrain the temperature and radius of the thermal emitter observed in the first days of the SN 2020bvc emission. We have also included the $g$ and $r$ bands {\it ZTF} observations that are closer in time to each SED epoch, in order to extend the UV and optical wavelength range from $\sim$2,000 $\AA$ to $\sim$7,000 $\AA$. We used {\it Swift} data observed at MJD = 58884.0 and MJD = 58887.3 (see also Table \ref{tab:App1}) that have been complemented with {\it ZTF} data observed on MJD = 58884.5 and MJD = 58887.5, respectively, see also Table \ref{tab:App2}. Both datasets were fit using a black body model and two absorption models which take into account the Galactic interstellar reddening ($E(B-V) = 0.039$ mag) and an additional model that corrects for the dust grain extinction in the host galaxy, with the $E(B-V)$ parameter fixed to the value of 0.01 mag, which was obtained from the Balmer decrement ratio analysis. For both models we used a \citet{Cardelli1989} extinction law. The fits were obtained using XSPEC \citep{Arnaud1996} and the results are shown in Table \ref{tab:App2a}, see also Fig.~\ref{fig:App1}. The normalization term is proportional to the luminosity of the thermal component, from which we estimated the radius $R$ of the emitter using the equation relating the luminosity from a black body with its radius and temperature $T$,
\begin{equation}
    L = 4 \pi R^2 \sigma T^4,
\end{equation}
with $\sigma$ being the Stefan--Boltzmann constant. We assume a distance to SN 2020bvc of $d = 120$ Mpc.

\begin{table}
\caption{Best-fit results obtained for the SEDs built using {\it Swift} and {\it ZTF} data on MJD = 58884.0 and MJD = 58887.3.}
\centering
\begin{tabular}{lccc}
\hline\hline
Epoch & kT & Norm. & Radius  \\
($MJD$) & ($eV$) & (10$^{-4}$ $erg/s/kpc$) & (10$^{15}$ $cm$) \\
\hline 
58884.0 & 1.23$^{+0.79}_{-0.44}$ & 4.15$^{+2.63}_{-1.89}$ & 1.42$\pm$0.56\\
58887.3 & 0.61$^{+0.38}_{-0.20}$  & 2.39$^{+1.32}_{-1.05}$ &  4.43$\pm$1.59\\
\hline
\end{tabular}
\label{tab:App2a}
\end{table}

\section*{C. Data Tables}

\begin{table}[h!]
\centering
\caption{{\it Swift}-UVOT data. The magnitudes are not corrected for Galactic extinction (E(B$-$V) = 0.01 mag).}
\begin{tabular}{lrrr}
\hline\hline
Filter & Epoch & magnitude & error \\
 & (MJD) & (mag) & (mag)\\
\hline
 UVOT-V  & 58884.0 & 17.18 & 0.22 \\
 UVOT-V  & 58885.6 & 17.12 & 0.22 \\
 UVOT-V  & 58886.5 & 17.47 & 0.26 \\
 UVOT-V  & 58887.3 & 17.24 & 0.25 \\
 UVOT-V  & 58888.9 & 17.13 & 0.22 \\
 UVOT-V  & 58891.6 & 16.59 & 0.14 \\
 UVOT-V  & 58893.5 & 16.40 & 0.14 \\
 UVOT-V  & 58895.0 & 16.25 & 0.17 \\
 UVOT-B  & 58884.0 & 17.22 & 0.15 \\
 UVOT-B  & 58885.6 & 17.45 & 0.17 \\
 UVOT-B  & 58886.5 & 17.63 & 0.18 \\
 UVOT-B  & 58887.3 & 17.48 & 0.18 \\
 UVOT-B  & 58888.9 & 17.02 & 0.14 \\
 UVOT-B  & 58891.6 & 16.93 & 0.12 \\
 UVOT-B  & 58893.5 & 17.01 & 0.13 \\
 UVOT-B  & 58895.0 & 17.34 & 0.17 \\
 UVOT-U  & 58884.0 & 17.12 & 0.14 \\
 UVOT-U  & 58885.6 & 18.27 & 0.21 \\
 UVOT-U  & 58886.5 & 18.18 & 0.2  \\
 UVOT-U  & 58887.3 & 17.96 & 0.19 \\
 UVOT-U  & 58888.9 & 18.19 & 0.2  \\
 UVOT-U  & 58891.6 & 18.03 & 0.15 \\
 UVOT-U  & 58893.5 & 18.20 & 0.17 \\
 UVOT-U  & 58895.0 & 18.22 & 0.21 \\
 UVOT-W1 & 58884.0 & 17.32 & 0.13 \\
 UVOT-W1 & 58885.6 & 19.08 & 0.24 \\
 UVOT-W1 & 58886.5 & 18.96 & 0.22 \\
 UVOT-W1 & 58887.3 & 19.25 & 0.26 \\
 UVOT-W1 & 58888.9 & 18.92 & 0.22 \\
 UVOT-W1 & 58891.6 & 19.25 & 0.19 \\
 UVOT-W1 & 58893.5 & 19.30 & 0.21 \\
 UVOT-W1 & 58895.0 & 19.34 & 0.28 \\
 UVOT-M2 & 58884.0 & 17.47 & 0.12 \\
 UVOT-M2 & 58885.6 & 19.80 & 0.25 \\
 UVOT-M2 & 58886.5 & 19.92 & 0.26 \\
 UVOT-M2 & 58887.3 & 19.86 & 0.27 \\
 UVOT-M2 & 58888.9 & 19.62 & 0.22 \\
 UVOT-M2 & 58891.6 & 20.17 & 0.22 \\
 UVOT-M2 & 58893.5 & 19.93 & 0.21 \\
 UVOT-M2 & 58895.0 & 19.71 & 0.26 \\
 UVOT-W2 & 58884.0 & 17.83 & 0.14 \\
 UVOT-W2 & 58885.6 & 19.64 & 0.26 \\
 UVOT-W2 & 58886.5 & 19.43 & 0.24 \\
 UVOT-W2 & 58887.3 & 19.82 & 0.29 \\
 UVOT-W2 & 58888.9 & 19.75 & 0.26 \\
 UVOT-W2 & 58891.6 & 19.73 & 0.21 \\
 UVOT-W2 & 58893.5 & 19.65 & 0.21 \\
 UVOT-W2 & 58895.0 & 20.39 & 0.37 \\
\hline
\end{tabular}
\label{tab:App1}
\end{table}

\begin{table}
\centering
\caption{{\it ZTF} \citep{Graham2019,Bellm2019} and ASAS-SN \citep{2014ApJ...788...48S} data. The magnitudes are not corrected for the Galactic extinction (E(B$-$V) = 0.01 mag).}
\begin{tabular}{lrrr}
\hline\hline
Filter & Epoch & magnitude & error \\
 & (MJD) & (mag) & (mag)\\
\hline
 ASAS-SN g & 58883.6 & >18.6 & - \\
 ASAS-SN g & 58884.1 & 17.0 & - \\
\hline
 ZTF-g & 58884.5 & 17.33 & 0.05 \\
 ZTF-g & 58887.5 & 17.06 & 0.05 \\
 ZTF-g & 58890.6 & 16.71 & 0.05 \\
 ZTF-g & 58893.5 & 16.64 & 0.04 \\
 ZTF-g & 58898.5 & 16.86 & 0.03 \\
 ZTF-g & 58903.5 & 17.18 & 0.05 \\
 ZTF-g & 58911.4 & 17.81 & 0.06 \\
 ZTF-g & 58911.6 & 17.77 & 0.08 \\
 ZTF-g & 58914.5 & 18.03 & 0.07 \\
 ZTF-g & 58936.5 & 18.91 & 0.12 \\
 ZTF-g & 58939.5 & 18.82 & 0.20 \\
 ZTF-g & 58943.4 & 18.80 & 0.12 \\
 ZTF-g & 58944.5 & 18.94 & 0.12 \\
 ZTF-r & 58884.5 & 17.55 & 0.06 \\
 ZTF-r & 58887.5 & 17.02 & 0.04 \\
 ZTF-r & 58890.5 & 16.59 & 0.04 \\
 ZTF-r & 58898.5 & 16.36 & 0.04 \\
 ZTF-r & 58903.5 & 16.48 & 0.04 \\
 ZTF-r & 58911.5 & 16.82 & 0.04 \\
 ZTF-r & 58914.5 & 17.07 & 0.04 \\
 ZTF-r & 58936.4 & 18.15 & 0.07 \\
 ZTF-r & 58939.4 & 18.13 & 0.09 \\
 ZTF-r & 58941.4 & 18.32 & 0.09 \\
 ZTF-r & 58943.5 & 18.32 & 0.08 \\
\hline
\end{tabular}
\label{tab:App2}
\end{table}

\begin{table}
\centering
\caption{{\it Swift}-XRT and {\it CXO} data. Fluxes and corresponding uncertainties are reported in columns 2 and 3 while the HR values, and its uncertainties, are in columns 4 and 5.}
\begin{tabular}{lcccc}
\hline\hline
Epoch & flux & error & HR & error \\
(MJD) & (erg/cm$^2$/s) & (erg/cm$^2$/s) & & \\
\hline 
58884.8 & <3.23e$-14$ & -- & -- \\
58886.9 & 1.20e$-13$  & 8.65e$-14$ & $-0.99$ & 0.71\\
58892.3 & 3.51e$-14$  & 2.19e$-14$ & $-0.93$ & 0.56 \\
\hline
58895.8 & 1.53e$-14$  & 0.54e$-14$ & $-0.39$ & 0.32 \\
58907.0 & 1.35e$-14$  & 0.51e$-14$ & $-0.77$ & 0.30 \\
\hline
\end{tabular}
\label{tab:App3}
\end{table}

\end{appendix}

\end{document}